\documentclass[twoside,openright,notitlepage,numbers=noenddot,headinclude,footinclude=true,cleardoublepage=empty,dottedtoc,BCOR=5mm,paper=a4,fontsize=11pt,]{article}

\usepackage{amsmath,amsfonts,amssymb,amsthm,bm,fancyhdr,color,graphicx,hyperref,subcaption,float,authblk,csquotes,titling,authblk,tocloft,blindtext,etoolbox}
\usepackage[usenames,dvipsnames,svgnames,table]{xcolor}
\usepackage[natbib=true,sorting=none,style=numeric-comp,backend=bibtex]{biblatex}

\DeclareMathOperator\arctanh{arctanh}

\newcommand\myshade{85}
\colorlet{mylinkcolor}{violet}
\colorlet{mycitecolor}{YellowOrange}
\colorlet{myurlcolor}{Aquamarine}

\hypersetup{
  linkcolor  = mylinkcolor!\myshade!black,
  citecolor  = mycitecolor!\myshade!black,
  urlcolor   = myurlcolor!\myshade!black,
  colorlinks = true,
}

\definecolor{myblue}{rgb}{.8, .8, 1}

\PassOptionsToPackage{
backend=bibtex8,bibencoding=ascii,
language=auto,
style=numeric-comp,
sorting=nyt, 
maxbibnames=10, 
natbib=true 
}{biblatex}

\newcounter{savecntr}

\begin{document}

\title{\textbf{Stability of Wilson loops and other observables in various type IIB Backgrounds}}
\author{Dimitrios Chatzis \setcounter{savecntr}{\value{footnote}}\thanks{dimitrios.chatzis@Swansea.ac.uk}}
\affil[1]{\small \centering{Department of Physics, Swansea University, Swansea SA2 8PP, United Kingdom}}
\date{}
    \maketitle

\begin{abstract}
	  This work involves the stability study of various observables (Wilson loop, 't Hooft loop and Entanglement Entropy) under linear fluctuations of the coordinates for certain ten-dimensional solutions of type $\mathrm{IIB}$ Supergravity that have appeared in the literature recently. These backgrounds are defined using intersecting and wrapped $\mathrm{D}5$ branes, their holographically dual field theories are speculated to confine and they have a non-local $\mathrm{UV}$ completion that is governed by Little String Theory. The present study confirms previous claims on the stability of the solutions made using the concavity condition for the energy of the probe strings, by studying the eigenvalue problem for each case and also providing a numerical analysis.

\end{abstract}

\newpage
\tableofcontents
\newpage

\section{Introduction}
Since the early days of the gauge/gravity correspondance \cite{Maldacena_1999,witten1998anti} and its generalisation to non-conformal settings with no $\mathrm{AdS}$ factor \cite{witten1998antide,Itzhaki_1998,Boonstra_1999,polchinski2000string,Girardello_2000,Klebanov_2000,MN,Kluson_2005,Kluso_2006,Kluso_2008,Kanitscheider_2008}, there have been many new insights into strong coupling phenomena of Quantum Field Theories. Various works have been done in this direction in the past, see \cite{Brandhuber_1998,kinar1998q,brandhuber1999wilson,sonnenschein2000,Herzog_2002,Klebanov_2006,Yamaguchi_2006,Nunez_2010} and many more, in order to study non-perturbative effects of some phenomenological resemblance to the theory of strong interactions. In particular, there has been a standard way of computing the Wilson loop, represented as the boundary of the worldsheet of a string exploring the holographic dimension \cite{Maldacena_1998}, through which one can examine the confinement of the dual theory using the usual arguments for the potential energy \cite{Wilson}. \\

It was realised that in some cases the classical probe string exploring the bulk, whose endpoints at radial infinity represent the static quark-anti-quark pair, could admit more than one embedding for a given set of boundary conditions. This allows for a phase transitions to occur between different classical configurations of the system, as the energy of the pair is a multivalued function of the length of separation. While there has been a criterion based on \cite{Bachas_1985,kinar1998q} for the stability of a given embedding that agrees with past examples, namely $\mathrm{d}^2E(L)/\mathrm{d}L^2\leq 0$ where $E(L)$ is the energy of the pair as a function of their separaration, the most trusted way of determining stability is through the study of small fluctuations around a classical solution.\\
 
In the current work, we will first present the four Type $\mathrm{IIB}$ Supergravity backgrounds of interest appearing in \cite{CMR1,CMR2}, and the behaviour of their dual field theories in Sections \ref{setup} and \ref{bgs}. These are ten-dimensional gravity backgrounds describing different setups of $\mathrm{D}5$ branes with corresponding dual theories of various dimensions, that have a non field-theoretic UV completion. Section \ref{observables} will concern the observables which we are going to study, as we will also be interested in 't Hooft loops and Entanglement Entropy, and review how they are calculated using holography. In Section \ref{pert} we will analyse the dynamics of small linear fluctuations and their eigenvalue problems, using techniques that were developed in \cite{SAS07,SAS08,SS08,Silva}. The goal of this study is to investigate if their spectrum includes negative eigenvalues for some range of values for the tip of the string, thus producing instabilities. Finally, the stability analysis will be carried out for backgrounds $\mathrm{I}-\mathrm{IV}$ in Section \ref{stabilitystudy}: The equations of motion for the perturbations are studied in the form of a Schr\"{o}dinger equation and their potentials are calculated in each case. We also include a numerical calculation of the eigenvalues when necessary, in order to further support our claims.

\section{The setup}\label{setup}
Consider the general non-diagonal metric of the form:

\begin{equation}\label{metric}
	\mathrm{d}\mathrm{s}^2 = G _{tt}\mathrm{d}t^2 + G _{x_ix_j}\mathrm{d}x^i \mathrm{d}x^j + G _{rr}\mathrm{d}r^2 + G _{x_ia}\mathrm{d}x^i \mathrm{d}\theta^a + G _{ab}\mathrm{d}\theta ^a \mathrm{d}\theta ^b,
\end{equation}

where $\{x^i\}$ are some spatial coordinates, $r$ is the holographic radial coordinate and $\{\theta^a\}$ are compact coordinates of the internal part of the space. All the metric components carry $r$-dependence and we let the components $G_{x_1 a}$ and $G_{ab}$ to depend on one of the internal coordinates $\theta$. The backgrounds that will be considered in this paper are part of ten-dimensional solutions of Type $\mathrm{II}$B supergravity. They will have spatial coordinates $x_1,x_2,\dots$ and a compact part consisting of spheres. More specifically, we will consider background $\mathrm{I}$  \cite{Khuri_1993,Cowdall_1998,Itzhaki_2006}, background $\mathrm{II}$ introduced in \cite{CMR1}, as well as backgrounds $\mathrm{III}$ and $\mathrm{IV}$ introduced in \cite{CMR2}. The first two solutions share a $\mathbb{R}^{1,1}$ Minkowski subspace and their geometry is of the form:

\begin{equation}
	\mathrm{I},\mathrm{II}:\,\,	\mathbb{R}^{1,1}\times \mathbb{R}_r\times \mathrm{S}^3\times\mathrm{S}^3\times \mathcal{M} _{\varphi},
\end{equation}

where $\mathcal{M} _{\varphi}=\mathbb{R}_{\varphi}$ and $\varphi \in \mathbb{R}$ for background $\mathrm{I}$, while $\mathcal{M} _{\varphi}=\mathrm{S}^1 _{\varphi}$ for background $\mathrm{II}$, in which case $\varphi$ is taken to be a compact cigar like direction that shrinks smoothly to a point $r _{+}$ where the space terminates. The other two backgrounds have a single three-sphere factor and the remaining 3 coordinates are either spanning real lines or a two sphere:

\begin{equation}
\begin{split}
	\mathrm{III}:& \,\, \mathbb{R}^{1,4}\times \mathrm{S}^1 _{\varphi}\times \mathbb{R} _r\times \mathrm{S}^3,\\
	\mathrm{IV}:& \,\, \mathbb{R}^{1,2}\times \mathrm{S}^1 _{\mu}\times \mathbb{R}_r \times \mathrm{S}^3\times \mathrm{S}^2,
\end{split}
\end{equation}

where again, $\varphi$ and $\mu$ express shrinking cycles. We will also define the following functions of the metric components for later convenience: 

\begin{equation}
f_{x_1}(r)=-G _{tt}G _{x_1x_1}, \,\, f _{ij}(r) = -G _{tt}G _{ij}, \end{equation}
\begin{equation}
	g(r) = -G _{tt}G _{rr},
\end{equation}
\begin{equation}
	h (r) = G _{rr}G_{x_1x_1}, 
\end{equation}
\begin{equation}
f _{ab}(r,\theta) = - G _{tt}G _{ab}, \,\, f _{ia}(r,\theta)=-G _{tt}G _{x_i a},
\end{equation}
Lastly, the backgrounds share another common feature, that is they all have logarithmic dilatons and thus the string coupling constants grow linearly for large values of the radial coordinate\footnote{ The coupling constants of the dual Quantum Field Theories were calculated in \cite{CMR1,CMR2}.}.

\section{Type $\mathrm{IIB}$ Supergravity solutions and their dual QFTs}\label{bgs}
\subsection{Background I}
We consider the following solution of Type $\mathrm{IIB}$ supergravity, introduced in \cite{Itzhaki_2006,Khuri_1993} describing the intersection of two stacks of $\mathrm{D}5$ branes along $\mathbb{R}^{1,1}[t,x]$, while each stack extends further on a four-dimensional space $\mathcal{M}_4\cong \mathbb{R}^4$. Considering the backreaction of the stacks at strong coupling forces the branes to also share the non-compact $\varphi$ direction, where each stack wraps around a three-sphere $\hat{\mathrm{S}}^3[\theta_A,\phi_A,\psi_A]$, $\tilde{\mathrm{S}}^3[\theta_B,\phi_B,\psi_B]$. After setting $\alpha ^{\prime}=1=g_s$ the metric is expressed in the string frame as:

\begin{equation}\label{BGI}
	\begin{split}
		\mathrm{d}\mathrm{s}^2 _{(\mathrm{I})} = r &\left\{ -\mathrm{d}t^2 + \mathrm{d}x^2 + \frac{(e_A^2 + e_B^2)}{2}\mathrm{d}\varphi ^2 + \frac{8}{r^2\left( e_A^2 + e_B^2 \right) }\mathrm{d}r^2\right. \\
		&\left.+ \frac{2}{e_A^2}\left( \hat{\omega}_1^2 +\hat{\omega}_2^2 +\hat{\omega}_3^2   \right) + \frac{2}{e_B^2}\left( \tilde{\omega}_1^2 +\tilde{\omega}_2^2 +\tilde{\omega}_3^2  \right)    \right\} ,
	\end{split}
\end{equation}
where $e_A$ and $e_B$ are fixed parameters and by $\hat{\omega}_i$ and $\tilde{\omega}_i$ we note the two sets of Maurer-Cartan $\mathrm{SU}(2)$ one-forms for the three-spheres $\mathrm{\hat{S}}^3,\mathrm{\tilde{S}}^3$ parametrized by $(\theta_A,\phi_A,\psi_A)$ and $(\theta_B,\phi_B,\psi_B)$ respectively\footnote{$\theta_{A,B}\in [0,\pi), \psi_{A,B}\in[0,4\pi],\phi_{A,B}\in[0,2\pi]$}:

\begin{equation}\label{MC}
\begin{matrix}
	\begin{split}
		&\hat{\omega}_1 = \cos\psi_A \mathrm{d}\theta_A + \sin\psi_A\sin\theta_A \mathrm{d}\phi_A ,\\
		&\hat{\omega}_2 = -\sin\psi_A \mathrm{d}\theta_A + \cos\psi_A \sin\theta_A \mathrm{d}\phi_A,\\
		&\hat{\omega}_3 = \mathrm{d}\psi_A + \cos\theta_A \mathrm{d}\phi_A,
	\end{split} &&  
	\begin{split}
		&\tilde{\omega}_1 = \cos\psi_B \mathrm{d}\theta_B + \sin\psi_B\sin\theta_B \mathrm{d}\phi_B ,\\
		&\tilde{\omega}_2 = -\sin\psi_B \mathrm{d}\theta_B + \cos\psi_B \sin\theta_B \mathrm{d}\phi_B,\\
		&\tilde{\omega}_3 = \mathrm{d}\psi_B + \cos\theta_B \mathrm{d}\phi_B,
	\end{split}
\end{matrix}
\end{equation}
 For completeness, we also present the Ramond-Ramond three-form, its two-form potential and the dilaton:

\begin{equation}\label{F3}
	F _{3}= \mathrm{d} C_2 = -\frac{2}{e_A^2}\hat{\omega}_1\wedge\hat{\omega}_2\wedge \hat{\omega}_3 -\frac{2}{e_B^2}\tilde{\omega}_1\wedge\tilde{\omega}_2\wedge \tilde{\omega}_3 ,
\end{equation}

\begin{equation}\label{C2}
C _{2}= -\frac{2}{e_A^2}\psi _{A}\sin\theta_A \mathrm{d}\theta_A\wedge \mathrm{d}\phi_A-\frac{2}{e_B^2}\psi _{B}\sin\theta_B \mathrm{d}\theta_B\wedge \mathrm{d}\phi_B,
\end{equation}

\begin{equation}\label{Phi}
\Phi = \log r\,\, , \,\, g_s \sim r.
\end{equation}

The number of branes of each stack is given by the flux of $F_3$ over the sphere the other stack is extending, which gives:
\begin{equation}
N_A = \frac{8}{e_B^2} \,\, , \,\, N_B = \frac{8}{e_A^2},
\end{equation}

and thus the parameters $e_A,e_B$ are quantized. \\

\textbf{Low energy behaviour:} As $r\to 0$ the background is singular (the Ricci scalar diverges) and the description using \eqref{BGI}-\eqref{Phi} is not trustable. In the IR, the dual field theory is described by a $(1+1)$-dimensional two node quiver gauge theory with gauge group $\mathrm{SU}(\mathrm{N}_\mathrm{A})\times \mathrm{SU}(\mathrm{N}_\mathrm{B})$, which also contains massless chiral fermions in the bifundamental\footnote{these belong to the spectrum as one can consider open strings with their endpoints constrained on the branes.} $(\textbf{N}_A,\overline{\textbf{N}}_B)$. We make note that the theory on the intersection is not independent of the worldvolume theory of the D-branes, as there is an inflow from the Chern-Simons terms from the bulk in order to cancel the anomalies on the intersection. \\

\textbf{High energy behaviour:} In the strong coupling regime of the theory, that is large values of the holographic coordinate $r$, the $\varphi$ coordinate is included to the effective theory on the intersection ($[t,x,\varphi]$) making it $(2+1)$-dimensional, while the amount of supersymmetry preserved also increases, see \cite{CMR1} for more details. For even higher values of the dilaton, which diverges in the far UV, in order to have a trustable description of the system one needs to use $\mathrm{S}$-duality, after which the system is described by $\mathrm{NS}5$ branes and two Little String Theories intersecting along the same coordinates as the branes do.\\

We therefore conclude that the field theory dual to the singular background \eqref{BGI}-\eqref{Phi}, while initially describing an effective $(1+1)$-dimensional theory on the intersection, as one approaches the strong coupling regime an extra dimension emerges and the theory becomes $(2+1)$-dimensional. While initially the solution preserves eight supercharges, the addition of a spatial dimension enhances the Poincar\'e group from $\mathrm{SO}(1,1)$ to $\mathrm{SO}(2,1)$ and the number of preserved supercharges grows to sixteen. Then for very large coupling the theory is UV completed by a non field theoretic description of Little String Theories.

\subsection{Background II}
We now turn to the second background of interest, introduced in \cite{CMR1}, which is to be thought of as a smooth version of background I, resolving the singularity at $r=0$ by compactifying $\varphi$ as a cigar-like coordinate on a smoothly shrinking circle and introducing a fibration of this circle and the three-spheres:

\begin{equation}\label{BGII}
	\begin{split}
		\mathrm{d}\mathrm{s}^2 _{(\mathrm{II})} = r &\left\{ - \mathrm{d}t^2 + \mathrm{d}x^2 + f _{s}(r) \mathrm{d}\varphi^2 + \frac{4}{r^2 f _{s}(r)}\mathrm{d}r^2\right.  \\
		&+ \frac{2}{e_A^2} \left[ \hat{\omega}_1^2 +\hat{\omega}_2^2 +\left( \hat{\omega}_3 - e_A Q_A \zeta(r)\mathrm{d}\varphi\right)^2    \right]\\
		& \left. +\frac{2}{e_B^2} \left[ \tilde{\omega}_1^2 +\tilde{\omega}_2^2 +\left( \tilde{\omega}_3 - e_B Q_B \zeta(r)\mathrm{d}\varphi\right)^2    \right]\right\},
	\end{split}
\end{equation}

where the functions appearing in the metric are:

\begin{equation}
\begin{split}
	&f _{s}(r) := \frac{e_A^2+e_B^2}{2} - \frac{m}{r^2} - \frac{2(Q_A^2+Q_B^2)}{r^4}\equiv \frac{e_A^2+e_B^2}{2r^4}\left( r^2-r _{+}^2 \right) \left( r^2- r _{-}^2 \right), \\
	&\zeta (r) := \frac{1}{r^2} - \frac{1}{r _{+}^2},\\
	&r _{\pm}^2 := \frac{m \pm \sqrt{m^2 + 4\left( Q_A^2+Q_B^2 \right) \left( e_A^2+e_B^2 \right) }}{e_A^2+e_B^2}.
\end{split}
\end{equation}

This background preseves four supercharges if the parameters satisfy $m=0$ and $e_AQ_B = \pm e_B Q_A$, while for $m=0=Q_A=Q_B$, background I is recovered. The rest of the fields read:

\begin{equation}
	\begin{split}
		F _{3}= \mathrm{d}C_2 =& 2\zeta ^{\prime}(r) \mathrm{d}r\wedge \mathrm{d}\varphi\wedge  \frac{Q_A}{e_A}\hat{\omega}_3 + \frac{2}{e_A^2}\hat{\omega}_1\wedge\hat{\omega}_2\wedge \left(e_AQ_A\zeta(r)\mathrm{d}\varphi - \hat{\omega}_3  \right) \\
		&+ (\hat{\omega}_i\leftrightarrow \tilde{\omega}_i, A \leftrightarrow B) 
	\end{split},
\end{equation}

\begin{equation}
	\begin{split}
		C_2 =& \psi _A \left( \frac{2Q_A}{e_A}\zeta ^{\prime}(r) \mathrm{d}r\wedge \mathrm{d}\varphi - \frac{2}{e_A^2}\sin\theta_A \mathrm{d}\theta_A\wedge \mathrm{d}\phi_A \right) + \frac{2}{e_a}\cos\theta_A Q_A \zeta(r) \mathrm{d}\varphi \wedge \mathrm{d}\phi _A\\
		&+ (\hat{\omega}_i\leftrightarrow \tilde{\omega}_i, A \leftrightarrow B) 
	\end{split},
\end{equation}

\begin{equation}
\Phi = \log r .
\end{equation}
In order for $\mathrm{S}^1[\varphi]$ to shrink smoothly at the point $r=r _{+}$, the period of $\varphi$ is fixed to be:

\begin{equation}
	L _{\varphi} = \frac{8\pi}{r _{+}f _s ^{\prime}(r _{+})}.
\end{equation}

\textbf{Low energy behaviour:} The choice of periodicity $L _{\varphi}$ forces the space to be smooth and free from conical singularities, as the Ricci scalar does not diverge in the range $[r _{+},\infty)$ and the dilaton is bounded from below. The role of $\varphi$ is now different, as it is compact, and as $r\to r _{+}$ (and $f_s(r)\approx 0$) the effective description is again a $(1+1)$-dimensional $\mathrm{SU}(\mathrm{N}_\mathrm{A})_{\mathrm{N}_\mathrm{B}}\times \mathrm{SU}(\mathrm{N}_\mathrm{B})_{\mathrm{N}_\mathrm{A}}$ gauge theory, that is dependent on the D-brane bulk theory as in the case of background $\mathrm{I}$. The compactification of $\varphi$ also results in the appearance of a massless Kaluza-Klein spectrum on the branes.\\

\textbf{High energy behaviour:} At higher values of the coupling, $\varphi$ decompactifies, thus the effective theory is $(2+1)$-dimensional, while for even higher energies the dual theory is again completed by the non local description of Little String Theories.\\

As was the case with the first background, the increase in spatial dimensions of the effective theory on the intersection for higher energies also increases the amount of supersymmetry preserved: While for lower energies this system preserves four supercharges, the addition of $\varphi$ increases this number to eight. The reader can review \cite{CMR1} for a detailed discussion on supersymmetry preservation of backgrounds $\mathrm{I}$ and $\mathrm{II}$.

\subsection{Background III}

We will now consider the following background that was recently introduced in \cite{CMR2} describing a single stack of $N$ $\mathrm{D}5$ branes wrapped around a circle $\mathrm{S}^1[\varphi]$, which is again shrinking smoothly to zero at the point $r=r _{+}$ as in \eqref{BGII}. This geometry also includes a fibration of $\mathrm{S}^1[\varphi]$ and $\mathrm{S}^3[\theta,\phi,\psi]$, but now we have only one three-sphere and 4 spatial directions\footnote{where $\mathrm{d}x^2 _{1,\mathrm{n}}$ is an abbreviation for the Minkowski metric $-\mathrm{d}t^2 + \mathrm{d}x_1 ^2 + \dots + \mathrm{d}x_\mathrm{n}^2$}:

\begin{equation}
	\resizebox{.91\hsize}{!}{$\mathrm{d}\mathrm{s}^2 _{(\mathrm{III})}=r \left\{ \mathrm{d}x^2 _{1,4}+ f_s(r)\mathrm{d}\varphi^2 + \frac{N \mathrm{d}r^2}{r^2f_s(r)} + \frac{N}{4}\left[ \omega_1^2 + \omega_2^2 + \left( \omega_3 - \sqrt{\frac{8}{N}} Q \zeta(r) \mathrm{d}\varphi \right) ^2  \right]  \right\} ,$}
\end{equation}

where $\omega _i$ are again the Maurer-Cartan one-forms of $\mathrm{SU}(2)$. Here, we define a rescaled version of $f_s$ from background II, such that $f_s(r)\to 1$  as $r\to \infty$:

\begin{equation}
	f_s(r) = 1 - \frac{m}{r^2} - \frac{2Q^2}{r^4} = \frac{(r^2-r _{+}^2)(r^2- r _{-}^2)}{r^4},
\end{equation}

\begin{equation}
2r _{\pm}^2 = m \pm \sqrt{m^2+8Q^2},
\end{equation}

and $\zeta(r)$ is as before. The period of $\varphi$ is now:

\begin{equation}
	L _{\varphi} = \frac{2\pi \sqrt{N}r _{+}^2}{(r _{+}^2 - r _{-}^2)}=\sqrt{N}\pi \left( 1 + \frac{m}{\sqrt{m^2+8Q^2}} \right) .
\end{equation}

\textbf{Low energy behaviour:} The space is smooth in the range $[r _{+},\infty)$ and the dual theory to this background is $(4+1)$-dimensional for small values of the holographic coordinate near $r _{+}$.\\

\textbf{High energy behaviour:} The behaviour is similar to background II, as the coordinate $\varphi$ is non compact ($f_s(r)\approx 1$) in the UV and the dual Quantum Field Theory is $(5+1)$-dimensional while for even higher energies the UV completion is achieved by moving to the NS5 brane setup, just as is the case with the previous geometries.\\

We note that the dual field theory for this gravity background preserves eight supercharges when the parameter $m=0$. The supersymmetry preservation mechanism is again interesting, as it uses the technique developed in \cite{Anabalon_2021} and can be read in \cite{CMR2} where a thorough analysis is given.

\subsection{Background IV}
The last background we work with in this paper, also introduced in \cite{CMR2}, describes a stack of $\mathrm{D}5$ branes wrapping a two-sphere in a non supersymmetry-preserving manner. The bosonic fields are\footnote{the solution also admits a six-form potential $\mathrm{C}_6$ which we will not write here explicitly (see \cite{CMR2}).}:

\begin{equation}
	\mathrm{d}\mathrm{s}^2 _{(\mathrm{IV})}= r \left\{ \mathrm{d}x^2 _{1,2} + \left( 1 - \frac{m}{r^2} \right) \mathrm{d}\mu^2 + \frac{2 \mathrm{d}r^2}{r^2 \left( 1- \frac{m}{r^2} \right) }+ \mathrm{d}\vartheta^2 + \sin ^2 \vartheta \mathrm{d}\varphi ^2+ \sum _{i=1} ^{3}\left( \Theta ^{i} \right) ^2  \right\} ,
\end{equation}

\begin{equation}
	F _3 = 2N \mathrm{Vol}(\mathrm{S}^3) + \frac{N}{4} \mathrm{d} \left[ \omega _1 \wedge A ^{(1)} + \omega_2\wedge A ^{(2)} + \omega_3 \wedge A^{(3)} \right] ,
\end{equation}

\begin{equation}
\Phi = \log \left( \frac{4r}{N} \right) ,
\end{equation}

where the coordinates $x_2,x_3$ are non compact, $\mu$ is expressing a shrinking $S^{1}[\mu]$ whose period is $L _{\mu}=\sqrt{8}\pi$, $\vartheta$ and $\varphi$ parametrize a two-sphere $S^2[\vartheta,\varphi]$ and there is also a three-sphere $S^3[\theta,\phi,\psi]$ that is expressed using the one-forms $\Theta ^{i}$. The definition of $\Theta ^{i}$ is done by introducing a gauge field $A$ as\footnote{This satisfies: $\left( A^1 \right) ^2 + \left( A^2 \right) ^2 + \left( A^3 \right) ^2 = \mathrm{d}\mathrm{s}^2 \left( \mathrm{S}^2[\vartheta,\varphi] \right)= \mathrm{d}\vartheta ^2 + \sin^2\vartheta \mathrm{d}\varphi^2$} :

\begin{equation}
\begin{split}
	& A ^{1}= \sin\vartheta\cos\vartheta \cos\varphi\mathrm{d}\varphi,\\
	&A ^{2}= -\cos\varphi \mathrm{d}\vartheta + \cos\vartheta\sin\vartheta\sin\varphi \mathrm{d}\varphi\,\, , \,\, A ^{3}= - \sin^2\vartheta \mathrm{d}\varphi.
\end{split}
\end{equation}

Then the $\Theta$ one-forms are defined as:

\begin{equation}
\Theta ^{1} = \omega _1 - A ^1\,\, , \,\, \Theta ^{2} = \omega_2 - A^2 \,\, ,\,\, \Theta ^3 = \omega_3 - A^3,
\end{equation}

where $\omega _i$ are the usual Maurer-Cartan forms of $\mathrm{SU}(2)$ defined in \eqref{MC}.\\

\textbf{Low energy behaviour:} The space of this background ends at $r=\sqrt{m}$ in the far $\mathrm{IR}$, where $\mathrm{S}^1[\mu]$ shrinks. The theory living on the $\mathrm{D}5$ branes extended in $[x_1,x_2,\mu,\vartheta,\varphi]$ is $(5+1)$-dimensional, but due to the compactification on $\mathrm{S}^2[\vartheta,\varphi]$ and the shrinking cycle, the dual theory is $(2+1)$-dimensional in low energies and for\footnote{The case $m=0$ is not trustable in this description, as both the Ricci scalar and dilaton diverge} $m>0$. As was noted in \cite{CMR2}, the two-sphere $\mathrm{S}^2[\vartheta,\varphi]$ is not shrinking, therefore this theory is really six-dimensional.\\

\textbf{High energy behaviour:} In large values of the radial coordinate, this background has the same behaviour as the previous one, with the difference being that three directions are compactified on a circle and a two-sphere. This theory is UV completed by Little String Theory as well.\\

\section{Observables of interest}\label{observables}
We will probe various objects to get the Wilson loop, 't Hooft loop and also the Entanglement Entropy for each case, and study the stability of the embeddings under small coordinate perturbations, in order to deduce if the observables have been calculated with the correct/stable way in \cite{CMR1,CMR2}.
\subsection{Wilson loops}
For the classical embedding of the string, we choose to fixing its end points at $x_1=\pm L/2$ and $r=\infty$ (interpreted as the probe quark and anti-quark) and also consider static solutions. Thus we parametrize the embedding as $\left( t(\tau), x_1(\sigma), r(\sigma) \right)$ and keep the rest of the coordinates constant. Then the induced metric, after using reparametrization invariance to let $t=\tau$ (static gauge), is written:

\begin{equation}
	\mathrm{d}\mathrm{s}^2 _{\mathrm{ind}}= G _{tt}\mathrm{d}\tau^2 + \left( G _{rr}r ^{\prime 2} + G _{x_1x_1}x_1 ^{\prime 2} \right) \mathrm{d}\sigma ^2,
\end{equation}

where the prime denotes the derivative with respect to $\sigma$. Then, the Nambu-Gotto action reads\footnote{where by $\mathcal{T}$ we denote the temporal extent of the loop.}:

\begin{equation}\label{S0}
	S = - \frac{\mathcal{T}}{2\pi}\int \mathrm{d}\sigma \sqrt{f _{x_1}(r) x_1 ^{\prime 2} + g(r)r ^{\prime 2}}= - \frac{\mathcal{T}}{2\pi}\int \sqrt{f _{x_1}(r) \mathrm{d}x_1^2 + g(r) \mathrm{d}r ^2}  .
\end{equation}

The quantity under the square root defines an effective two-dimensional space on which geodesics correspond to Wilson loop configurations. The Euler-Lagrange equation for $x_1$ is:
\begin{equation}\label{eqemb}
	\frac{f _{x_1}(r) x _1 ^{\prime}}{\sqrt{f _{x_1}(r) x _1 ^{\prime 2} + g(r) r ^{\prime 2}}}   = \mathrm{const.} \equiv \sqrt{f _{x_1}(r_0)} 
\end{equation}

\begin{equation}
\Rightarrow	x_1 ^{\prime}(\sigma)=\pm \sqrt{\frac{g(r) f _{x_1}(r_0)}{f _{x_1}(r) \left[ f _{x_1}(r) - f _{x_1}(r_0)  \right] }}	r ^{\prime }(\sigma),
\end{equation}

and using reparametrization invariance we can choose $\sigma = r$, so the static solution reads:

\begin{equation}\label{xcl}
	x _{1\mathrm{cl}}^{\prime}(r)= \pm \frac{\sqrt{f_{x_1}(r_0)F(r)}}{f_{x_1}(r)}.
\end{equation}
	where $F$ is defined to be\footnote{It is apparent that the expression under the square root that appears in the Lagrangian is exactly $F$. Using \eqref{xcl}: $f _{x_1}(r)x _{1\mathrm{cl}} ^{\prime 2} + g(r) = \frac{f _{x_{1}}(r_0)g(r)}{f _{x_1}(r) - f _{x_{1}}(r_0)} + g(r) = \frac{f _{x_1}(r)g(r)}{f _{x_1}(r) - f _{x_{1}}(r_0)}= F(r)$} :
\begin{equation}\label{F}
	F(r):=\frac{f_{x_1}(r)g(r)}{f_{x_1}(r)-f_{x_1}(r_0)}
\end{equation}
and $r_0$ is the turning point of the string\footnote{In our parametrization, this satisfies $x_{1\mathrm{cl}} ^{\prime}(r_0)=\infty$} where the two branches of the solution \eqref{xcl} join. Then by integrating \eqref{xcl}, one gets the separation $L$ of the quark anti-quark pair:

\begin{equation}\label{length}
	L(r_0) = 2 \sqrt{f _{x_1}(r_0)}\int _{r_0}^{\infty}\mathrm{d}r \frac{ \sqrt{F(r)}}{f _{x_1}(r)}.
\end{equation}

Also, plugging the classical solution into \eqref{S0}, we can read off the expression for the potential energy of the system (using that $S\sim T\cdot E$), which we need to regularize by subtracting the infinite contribution due to the masses of strings extending from $\pm L/2$ at $r=\infty$ down to the point $r^{\star}$ where the space terminates:

\begin{equation}\label{energy}
	E(r_0) = \frac{1}{\pi}\int _{r_0}^{\infty}\mathrm{d}r \sqrt{F(r)} - \frac{1}{\pi}\int _{r^{\star}}^{\infty}\mathrm{d}r \sqrt{g(r)}
\end{equation}

We need to make note of the two conditions the energy as a function of the length must satisfy, that is, to produce an attractive force:

\begin{equation}\label{dEdL}
	\frac{\mathrm{d}E(L)}{\mathrm{d}L}=\frac{\mathrm{d}E(L)}{\mathrm{d}L}=\frac{\mathrm{d}E(r_0)}{\mathrm{d}r_0} \frac{1}{L^{\prime}(r_0)}=\frac{\sqrt{f _{x_1}(r_0)}}{2\pi}>0,
\end{equation}

which holds for every Supergravity background we are interested in ($f^{\prime}_{x_1}(r_0)>0$), and also the concavity condition \cite{Bachas_1985}:

\begin{equation}\label{concavity}
	\frac{\mathrm{d}^2E(L)}{\mathrm{d}L^2}=\frac{1}{L^{\prime}(r_0)}\frac{\mathrm{d}}{\mathrm{d}r_0}\left( \frac{\sqrt{f _{x1}(r_0)}}{2\pi} \right) = \frac{f^{\prime}_{x1}(r_0)}{4\pi \sqrt{f _{x}(r_0)} L ^{\prime}(r_0)}\leq0,
\end{equation}
	namely, to be a non increasing function of the quark-anti-quark distance. We also know $\sqrt{f ^{\prime}_{x1}(r_0)}>0$ to be true for all backgrounds, therefore this condition is violated in the (potential) regions in which $L ^{\prime}(r_0)>0$. These are unphysical regions in which one might find the configuration of the string to be unstable under perturbations.\\

Finally, we can rewrite the energy to find an expression which also contains the length:
\begin{equation}
	\begin{split}
		E(r_0) = \frac{1}{\pi}&\int _{r_0}^{\infty}\mathrm{d}r \sqrt{\frac{g(r)}{f _{x_1}(r)}}\left( \frac{f _{x_1}(r_0)}{\sqrt{f _{x_1}(r)-f _{x_1}(r_0)}} + \sqrt{f _{x_1}(r)- f _{x_1}(r_0)} \right) \\
		&- \frac{1}{\pi}\int _{r ^{\star}}^{\infty}\mathrm{d}r \sqrt{g(r)}\\
	\end{split}
\end{equation}

	\begin{equation}\label{eqE}
	\Rightarrow E(r_0) = \frac{\sqrt{f _{x_1}(r_0)}}{2\pi}L(r_0) + \frac{1}{\pi}\int _{r_0} ^{\infty} \mathrm{d}r \sqrt{\frac{g(r) (f _{x_1}(r)- f _{x_1}(r_0))}{f _{x_1}(r)}} - \frac{1}{\pi}\int _{r ^{\star}}^{\infty}\mathrm{d}r\sqrt{g(r)}
\end{equation}

	In the large $L$ limit when the turning point of the U-shaped string explores the bulk deeply, reaching the end of the space $r_0\to r^{\star}$, the difference of the two integrals in \eqref{eqE} approaches zero and the potential is proportional to the length. For backgrounds which  $f _{x_1}(r^{\star})=0$, we can already see from \eqref{eqemb} that the differential equation for the string embedding changes its form and admits constant solutions, corresponding to the disconnected configuration of two straight strings extending from $r=\infty$ to the end of the space whose endpoints are moving in $x_1$ with zero energy. We remind the reader that the function appearing in \eqref{eqE} that defines the slope of the energy with respect to the length is interpreted as the effective tension of the dual chromoelectric string \cite{kinar1998q}:

	\begin{equation}
		T _{\mathrm{eff}} \equiv \frac{\mathrm{d}E(L)}{\mathrm{d}L}= \sqrt{f _{x_1}(r ^{\star})},
	\end{equation}

	which gives a rather physical argument about confinement and screening: If $f _{x_1}(r ^{\star})\neq 0$ there exist a non-zero tension and the system exibits confinement, while for a zero tension system, the quark and anti-quark can move away from each other at no energy cost, that is, the potential is screened. We also comment that we expect all of the Wilson loop configurations in this paper to be stable, since the concavity condition seems to be satisfied in all cases and moreover, the length as a function of $r$ is monotonically decreasing and thus there is no possibility of $E(L)$ to be multivalued \cite{kinar1998q}. That being said, the study of small fluctuations around a certain classical configuration is the most trusted method to reach conclusions about stability.

\subsection{'t Hooft loops}\label{thooft}
In order to study the 't Hooft loop, that is the magnetic analogue of the Wilson loop, we probe a $\mathrm{D}5$ brane extending at least in the $\mathbb{R}^{1,1}[t,x_1]$ Minkowski subspace in order to obtain the effective string, while the remaining four directions can extend into compact or non compact coordinates depending on the background. After the induced metric on the $\mathrm{D}5$ brane is inserted in the Dirac-Born-Infeld action, and after computing the integral on the internal space, one can reduce the expression to a similar integral as in the Wilson loop case:

\begin{equation}
	S _{\hat{\mathrm{D}}5} \left[x(r)  \right] = \int \mathrm{d}^6\sigma e ^{-\Phi} \sqrt{-\mathrm{det}(g _{\mathrm{ind}})}= \mathrm{A}\int _{r_0} ^{\infty}\mathrm{d}r \sqrt{f^{\mathrm{t}} _{x_1}(r) x ^{\prime 2}(r)  + g^{\mathrm{t}}(r) },
\end{equation}
where the constant $\mathrm{A}$ is dependent on the details of the background. Moreover, the functions $f ^{\mathrm{t}} _{x_1}(r)$ and $g ^{\mathrm{t}}(r)$ are read directly through the above expression for each case and are no longer defined using the metric components. Nevertheless, $T _{\mathrm{eff}}=\sqrt{f ^{\mathrm{t}}_{x_1}(r ^{\star})}$ is still interpreted as the effective tension of the chromomagnetic string. In the examples of backgrounds I-IV the 't Hooft loop configurations are expected to be unstable as $L^{\prime}(r_0)>0$ and $T _{\mathrm{eff}}=0$, signifying that the monopole-anti-monopole pair screens: It is preferable for the system to transition to the disconnected configuration of the two strings extending in the entirety of the bulk. This is a typical behaviour of the magnetic analogue of the Wilson line for confining systems.

\subsection{Entanglement Entropy}
For the definition of Entanglement Entropy we consider two entangled regions with a minimal codimension 2 manifold connecting them. We take one of the regions to be the strip of width $L _{\mathrm{EE}}$ and the remaining space to be the other one and define the action for the Entanglement Entropy:

\begin{equation}\label{SEE}
	S _{\mathrm{EE}} = \frac{1}{4G_N}\int _{\Sigma _8}\mathrm{d}^{8}\sigma \sqrt{e^{-4\Phi}\mathrm{det}(g _{\Sigma _8})}
\end{equation}

As with the Wilson and 't Hooft loops, one can have a phase transition between different embeddings (8-manifolds in this case). Moreover, there is evidence to support that a phase transition in Entanglement Entropy signals confinement of the dual theory and therefore this observable can serve as an order parameter for confinement \cite{Klebanov_2008,Ryu_2006,Nunez_2014,Jokela_2021}. As was pointed out in \cite{Nunez_2014,Jokela_2021,CMR1}, there is a subtlety for theories that admit non-local $\mathrm{UV}$ completions: In such cases one must introduce a $\mathrm{UV}$ cutoff in order to make the phase transition apparent. The 8-manifolds that we are going to embed are extending either in four non compact spatial directions and four compact internal coordinates, or two spatial and six compact ones (in the case of background II where there are two three-spheres). After inducing the metric on the 8-manifold in the aforementioned gauge, the action \eqref{SEE} can also be brought, after computing the integrals, to the usual integral form:

\begin{equation}
	S _{\mathrm{EE}} = \mathrm{B} \int _{r_0} ^{\infty} \mathrm{d}r \sqrt{f ^{\mathrm{EE}} _{x_1}(r) x^{\prime 2} (r) + g ^{\mathrm{EE}} (r)},
\end{equation}

again as in the 't Hooft loop case, the functions $f ^{\mathrm{EE}} _{x_1}$ and $g ^{\mathrm{EE}}$ are to be read for the expression of each background separately and the constant $\mathrm{B}$ in front of the integral depends on the background.

\section{Perturbative study}\label{pert}
In this section we are going to present the derivation of the equations of motion that control the dynamics of small linear perturbations of the spacetime coordinates. In particular, we will derive them in the form of a Sturm-Liouville problem and study the eigenvalues for each case, having the turning point $r_0$ as a parameter.
\subsection{Perturbations and their equations of motion}
The perturbative study will be done in the spirit of \cite{SAS07,SAS08,SS08,Silva} by considering small fluctuations in the coordinates of \eqref{metric} as:

\begin{equation}
	\left.\Big(t, x_1,x_i,r,\theta _a \right.\Big) = \left.\Big(t, x _{1\mathrm{cl}}(r) + \delta x_1(r,t), \delta x_i(r,t),r,\theta _{a 0} +\delta \theta _a (r,t) \right.\Big) 
\end{equation}

This leads to an expansion of the action with respect to the perturbations, where the $\mathcal{O}(\delta x^{\mu 0})$ term is just \eqref{S0}:

\begin{equation}
	\mathcal{O}(\delta x^{\mu 1}): \,\,S_1 = - \frac{1}{2\pi}\int \mathrm{d}t \mathrm{d}r  \frac{f_{x_1} x_1^{\prime}}{F^{1/2}}\delta x_1^{\prime} , 
\end{equation}

\begin{equation}\label{act}
	\begin{split}
		\mathcal{O}(\delta x^{\mu 2}):S_2 = - \frac{1}{2\pi}\int \mathrm{d}t \mathrm{d}r &\left\{ \frac{f _{x_1}g}{2F ^{3/2}}\delta x_1^{\prime 2} - \frac{h}{2F ^{1/2}}\delta \dot{x}_1^{2}+\frac{f _{ij}}{2F ^{1/2}}\delta x_i^{\prime }\delta x_j^{\prime }\right. \\
		& - \frac{f _{ij}h F^{1/2}}{2f _{x_1}g}\delta \dot{x}_i\delta \dot{x}_j +  \frac{f _{ab}}{2F ^{1/2}}\delta \theta _a^{\prime }\theta _b ^{\prime }\\
		&\left. - \frac{f _{ab}hF ^{1/2}}{2f _{x_1}g}\delta \dot{\theta}_{a}\delta \dot{\theta}_{b}- \frac{f _{ia}h F^{1/2}}{g f _{x_1}}\delta \dot{x}_i\delta \dot{\theta}_{a}\right\}.
	\end{split}
\end{equation}

The first order contribution, using \eqref{xcl}, is a total derivative and thus can be omitted\footnote{\[S_1 = - \frac{1}{2\pi}\int \mathrm{d}t \mathrm{d}r \frac{f _{x_1}(r)x_{1\mathrm{cl}}^{\prime}(r) }{F^{1/2}(r)}\delta x_1^{\prime}(r,t)= \mp \frac{\sqrt{f _{x_1}(r_0)}}{2\pi}\int \mathrm{d}t \mathrm{d}r \frac{\mathrm{d}}{\mathrm{d}r}\delta x_1(r,t).\]One can expect this to be true in general, since we consider first order corrections of the action around a classical solution of the equation of motion.}. The Lagrangian at second order, when imposing harmonic time dependence on the perturbations: $\delta x^{\mu}(r,t) = e^{-i\omega t}\delta x^{\mu}(r)$, is of the form:

\begin{equation}
	\mathcal{L} ^{(2)}= \frac{1}{2}\Omega _{ij} (r)\delta \Phi ^{\prime i}\delta \Phi ^{\prime j} - \frac{\omega^2}{2}M _{ij}(r)\delta \Phi ^{i}\delta\Phi ^{j},
\end{equation}

where $\delta \Phi^i =(\delta x_1, \delta x_i,\delta \theta_a)^{\intercal}$. The equations of motion for the system are:

\begin{equation}\label{EOMmatrix}
\frac{\mathrm{d}}{\mathrm{d}r}\left( \Omega _{ij}\delta \Phi ^{\prime j} \right) =\omega^2 M _{ij}\delta \Phi ^j ,
\end{equation}

where the matrices $\Omega$ and $M$ have in general non zero non-diagonal elements, that is the perturbations couple to each other. This fact together with the $r$-dependence of the matrix $\Omega$ makes the system difficult to diagonalize. One can usually consider the simpler problem of zero mode solutions ($\omega^2=0$) where we can write:

\begin{equation}\label{zm}
	\Omega _{ij}(r)\delta \Phi ^{ (0) \prime j} = c _{i}
	\Rightarrow \delta \Phi ^{(0) i} = c _j \int _{r_0}^{r} \mathrm{d}\xi \Omega ^{-1} _{ij}(\xi) + \tilde{c}_i,
\end{equation}

for some constants $c_i,\tilde{c}_i$. The existence of zero modes signifies the boundary between solutions with $\omega^2(r_0)>0$ (stability) and $\omega^2(r_0)<0$ (instability) \cite{SAS07}, the goal being to find if there exists a range of $r_0$ values for which the later is true. \\

The fact that the zero mode solutions can be written in the integral form \eqref{zm} would not be possible in the case where the metric components $G _{tt}$, $G _{x_1x_1}$ also had explicit dependence on a certain non-cyclic coordinate, eg an angular coordinate as happens in other backgrounds\footnote{see \cite{SAS07} for interesting examples where the orientation in the internal space produces an angular instability.}. This happens due to their expansions producing terms $\propto \delta \Phi$ that do not originate from time derivatives, and thus have no factor of $\omega^2$ in the equations of motion and would survive in the zero mode problem. This type of angular perturbations as well as the transverse ones, are coupled among themselves due to the quadratic in $\mathrm{d}x_i$ and $\mathrm{d}\theta _{a}$ terms. In the case of the metric introduced in \eqref{metric} representing all four gravity backgrounds, the nonzero $G _{x_i a}$ forces $\delta x_i$ and $\delta \theta _a$ to couple, while in the case where $G _{t x_i}\neq 0$ and also in the presence of a nonzero $B$ field, the longitudinal, transverse and angular perturbations would all couple to each other, see \cite{SAS08}. The equations of motion for our cases of interest can be written in the following form from \eqref{act}:

\begin{equation}\label{EOM}
	\left\{\frac{\mathrm{d}}{\mathrm{d}r}\left[\begin{pmatrix}
	\frac{f _{x_1}g}{F^{3/2}} && 0 && 0 \\	
	0 && \frac{f _{ij}}{ \sqrt{F}} && 0\\
	0 && 0 && \frac{f _{ab}}{ \sqrt{F}}
		\end{pmatrix} \frac{\mathrm{d}}{\mathrm{d}r} \right] +\omega^2 \begin{pmatrix} \frac{h}{ \sqrt{F}} && 0 && 0 \\ 0 && \frac{f _{ij}h \sqrt{F}}{f _{x_1}g}&& \frac{f _{ib}h \sqrt{F}}{f _{x_1}g}\\ 0 && \frac{f _{ij}h \sqrt{F}}{f _{x_1}g} && \frac{f _{ab}h \sqrt{F}}{f _{x_1}g}\end{pmatrix} \right\} \begin{pmatrix}\delta x_1 \\ \delta x_j \\ \delta \theta _b\end{pmatrix}=0,
\end{equation}

	where $r_0\leq r < \infty$ and $r_0\geq r^{\star}$. We can see that there are no real $"$angular$"$ perturbations, as the equations for $\delta \theta_a$ are of the same form as the $\delta x_j$ ($j\neq1$).\\

	It is in general more illuminating to use a change of variables to map \eqref{EOM} to a Schr\"{o}dinger equation, as it captures the information about the spectrum of the eigenvalues best. This formulation of the problem is reviewed in Appendix \ref{SCH}, however for convenience, we quote the main equations here. Considering the Sturm-Liouville problem for the longitudinal fluctuations $\delta x_1$, although the same technique holds for any decoupled fluctuation, that is read from \eqref{EOM} as:

\begin{equation}\label{SL}
	\begin{split}
		- \frac{\mathrm{d}}{\mathrm{d}r}&\left[ P(r;r_0) \frac{\mathrm{d}}{\mathrm{d}r}  \right] \delta x_1 (r) = \omega ^2 W(r;r_0) \delta x_1(r) \,\, ,\,\, r ^{\star} \leq r_0\leq r<\infty,\\
		&		P(r;r_0):= \frac{f _{x_1}(r)g(r)}{F ^{3/2}(r;r_0)}\,\, , \,\, W(r;r_0):=\frac{h(r)}{\sqrt{F(r;r_0)}},
	\end{split}
\end{equation}
	
	we can get a Schr\"{o}dinger equation for $\Psi(y):=(PW)^{1/4}(r;r_0)\delta x_1(r)$ that has the usual form\footnote{where it is understood that $y=y(r;r_0)$, as defined in the Appendix.}

\begin{equation}\label{SE}
	\left[ - \frac{\mathrm{d}^2}{\mathrm{d}y^2} + V(y,r_0)  \right] \Psi (y) = \omega ^2 \Psi (y)\,\, ,\,\, 0\leq y \leq y_0  .
\end{equation}

The potential, which we are going to study for each case, is given by the expression:

\begin{equation}\label{V(r)}
		V (r,r_0) =  \frac{P^{1/4} }{W ^{3/4}} \frac{\mathrm{d}}{\mathrm{d}r} \left[ \left( \frac{P }{W } \right) ^{1/2} \frac{\mathrm{d}}{\mathrm{d}r}\left( P W  \right) ^{1/4}  \right] .
\end{equation}

	Two final important remarks on the types of fluctuations, that hold for any type of metric, are the following: The transverse ones do not admit zero modes\footnote{Proof of this can be found in section 3.3.1 of \cite{SAS07}.} and the longitudinal zero modes are in one-to-one correspondance with the critical points of the length function (see Appendix \ref{zeromodes} for a proof of this). Since in our cases the fluctuations of $x_j$ and $\theta _a$ are all treated as transverse, do not coupled to the longitudinal ones and also their lowest eigenvalue is found to be positive, they will not be able to produce instabilities.

\subsection{Boundary conditions}
	We will now state the boundary conditions needed in order to formulate the Sturm-Liouville problem. As was described in previous works \cite{SAS07,SAS08,Pufu,Silva}, one needs to consider two boundary conditions: One at the endpoints of the string and one at the tip of the $\mathrm{U}$-shaped embedding at $r=r_0$. In order for the endpoints to stay fixed and also to obtain normalizable solutions, we impose the vanishing of the perturbations at $r=\infty$:

\begin{equation}
\left.	\delta x ^{\mu} (r) \right| _{r=\infty}=0.
\end{equation}

The second boundary condition is more subtle and forces the two branches of \eqref{xcl} to glue together at the turning point, which in the $r$-gauge is a singular point of the differential equation \eqref{EOM}. The equations of motion near $r=r_0$ take the form:

\begin{equation}
	\frac{\mathrm{d}}{\mathrm{d}r}\left[ (r-r_0)^{3/2} \frac{\mathrm{d}\delta x_1(r)}{\mathrm{d}r}  \right] + \frac{\omega ^2 h(r_0)}{f^{\prime}_{x_1}(r_0)}\sqrt{r-r_0}\delta x_1(r) = 0 ,
\end{equation}

\begin{equation}
	\frac{\mathrm{d}}{\mathrm{d}r}\left[ \sqrt{r-r_0}\frac{\mathrm{d}\delta \theta_a(r)}{\mathrm{d}r} \right] + \frac{\omega^2h(r_0)}{f^{\prime}_{x_1}(r_0)} \frac{\delta x_i(r)}{\sqrt{r-r_0}}=0,
\end{equation}

\begin{equation}
	\frac{\mathrm{d}}{\mathrm{d}r}\left[ \sqrt{r-r_0}\frac{\mathrm{d}\delta x_i(r)}{\mathrm{d}r} \right] + \frac{\omega^2h(r_0)}{f^{\prime}_{x_1}(r_0)} \frac{\delta \theta_a(r)}{\sqrt{r-r_0}}=0,
\end{equation}

	where again, we emphasize that for the metric \eqref{metric} the angular perturbations obey the same equations of motion as the transverse ones and therefore will be treated as such, that is, they do not have any zero modes as well. The solutions of the above have the following form near $r_0$, for some constants $\alpha _{0}, \alpha_1, \beta_0, \beta_1, \gamma_0, \gamma_1$:

\begin{equation}\label{deltaxsmall}
\delta x_1(r) \approx \alpha_0 + \frac{\alpha_1}{\sqrt{r-r_0}},
\end{equation}

\begin{equation}
\delta x_m(r) \approx \beta_0 + \beta_1 \sqrt{r-r_0},
\end{equation}

\begin{equation}
	\delta \theta(r) \approx \gamma_0 + \gamma_1 \sqrt{r-r_0}.
\end{equation}

One can see that $\delta x_1(t,r_0)$ is not physical at $r=r_0$ in this gauge, which is further supported by the fact that this fluctuation is oriented parallel to the worldvolume of the string, while $\delta r(t,x)$ is transverse to it and thus finite at $r_0$. The way one overcomes the singularity of the perturbation at this point and the way to do it is to define the variable:

 \begin{equation}
	 u:= r + \Delta (t,r)\,\, , \,\, \Delta (t,r) = \frac{\delta x_1(r,t)}{x ^{\prime} _{\mathrm{cl}(r)}}.
 \end{equation}
 
Then, one can see that the classical solution remains unperturbed:

\begin{equation}
	\begin{split}
		x _1 (r) &= x _{1\mathrm{cl}}(r) + \delta x_1(r,t) = x _{1\mathrm{cl}}(u-\Delta (r,t)) + \delta x_1(r)\\
		&\approx x _{1\mathrm{cl}}(u) - x ^{\prime} _{1\mathrm{cl}}(u)\Delta (r,t) + \delta x_1(r,t)\\
		&=x _{1\mathrm{cl}}(u) - x ^{\prime} _{1\mathrm{cl}}(r + \Delta (r,t))\Delta (r,t) + \delta x_1(r,t)\\
		&\approx x_{1\mathrm{cl}}(u) - x ^{\prime} _{1\mathrm{cl}}(r) \frac{\delta x_1(r,t)}{x ^{\prime} _{1\mathrm{cl}}(r)} + \delta x_1(r,t) = x _{1\mathrm{cl}}(u) .
	\end{split}
\end{equation}

We can also write,

\begin{equation}
	r= u - \Delta (r,t) = u - \Delta (u-\Delta(r,t),t)\approx u - \Delta(u,t) = u - \frac{\delta x_1(u,t)}{x ^{\prime} _{1\mathrm{cl}}(u)},
\end{equation}

to see that the physical $\delta r (u,t)$ perturbation at $u\to r_0$ is finite\footnote{the first approximate equation gives us the interpretation of $\Delta (u,t)=\delta r(u,t)$.}:

\begin{equation}
	r \approx r_0 - \frac{\delta x_1(u,t)}{x ^{\prime} _{1\mathrm{cl}}(u)} = r_0 - C \left( \alpha_0 + \frac{\alpha_1}{\sqrt{u-r_0}} \right) \sqrt{u-r_0}
\end{equation}

\begin{equation}
\Rightarrow\delta r \approx - C \left( C_0 \sqrt{u-r_0} + C_1 \right) ,
\end{equation}

where $C$ is a possible constant in front of \eqref{deltaxsmall}. Then, we can write the boundary condition 

\begin{equation}
	\left. \frac{\mathrm{d}\delta r(r,t) }{\mathrm{d}x_1}\right| _{r\to r_0} = 0,
\end{equation}

 which leads to:

 \begin{equation}
	 \left( \frac{\delta x ^{\prime}_1}{x ^{\prime} _{1\mathrm{cl}}} - \frac{\delta x_1}{x ^{\prime 2} _{1\mathrm{cl}}}x ^{\prime \prime} _{1\mathrm{cl}} \right) \frac{1}{x^{\prime} _{1\mathrm{cl}}}=0
 \end{equation}

 \begin{equation}\label{BCdx1}
\Rightarrow	 \delta x_1(r) + 2\left( r-r_0 \right) \frac{\mathrm{d}\delta x_1(r)}{\mathrm{d}r} = 0 \,\,\, \,, \mathrm{as} \,\,r\to r_0.
 \end{equation}

\section{Stability analysis for the geometries of interest}\label{stabilitystudy}
Having presented the study of small fluctuations of the background, we are ready to apply this to the solutions presented in Section \ref{pert} and examine the stability of the embeddings of the observables displayed in Section \ref{observables}. We will calculate the Schr\"{o}dinger potential in each case and also provide numerical results to back up the arguments when necessary.

\subsection{Backgrounds dual to (1+1)-dimensional QFTs}
We will categorize the different backgrounds with respect to the-dimensionality of their dual theories, in order to stress the fact that even though we are dealing with theories that do not even have the same number of dimensions, as far as confinement is concerned they have similar behaviours. In the present subsection we deal with the first two backgrounds, dual to two-dimensional theories.
\subsubsection{Wilson Loop}
The functions of the metric components for the first background read:

\begin{equation}
\begin{split}
	&f ^{(\mathrm{I})}_x(r) = -G_{tt} G _{xx}= r^2 \,\, , \,\, f _{\varphi} ^{(\mathrm{I})} (r)= -G _{tt}G _{\varphi\varphi}= \frac{r^2\left( e_A^2+e_B^2 \right) }{2}\\
	&g ^{(\mathrm{I})} (r)= -G_{tt}G_{rr} = \frac{8}{e_A^2+e_B^2} \,\, , \,\, h ^{(\mathrm{I})} (r)= G_{rr}G_{xx} = g ^{(\mathrm{I})}(r), \\
\end{split}
\end{equation}

and the classical solution can be computed form \eqref{xcl}:

\begin{equation}\label{xclI}
	x _{\mathrm{cl}}^{(\mathrm{II})\prime 2}(r)= \frac{8r_0^2}{r^2\left( r^2-r_0^2 \right) \left( e_A^2+e_B^2 \right) } .
\end{equation}

From this, the length of separation is found to be:
\begin{equation}\label{LI}
	L^{(\mathrm{I})}(r_0) = 2r_0 \sqrt{\frac{8}{e_A^2+e_B^2}}\int _{r_0} ^{\infty} \frac{\mathrm{d}r}{r\sqrt{r^2-r_0^2}}= \sqrt{\frac{8}{e_A^2+e_B^2}}\pi\equiv L ^{(\mathrm{I})} _{\mathrm{LST}},
\end{equation}
where we emphasize the fact that it is constant and independent of $r_0$, which is a particularly interesting result, as even the concavity condition cannot help as an indication for stability ($L ^{(\mathrm{I})\prime}(r_0)=0$). This is interpreted as the characteristic length of the Little String Theory that UV completes the dual Quantum Field theory and suggests there exist only one configuration for the system. The energy is from \eqref{energy}:

\begin{equation}\label{EI}
	\begin{split}
		E^{(\mathrm{I})}(r_0) = \frac{1}{\pi} \sqrt{\frac{8}{e_A^2+e_B^2}}\int _{r_0} ^{\infty} \frac{\mathrm{d}r r^2}{\sqrt{r^2-r_0^2}} - \frac{1}{\pi}\sqrt{\frac{8}{e_A^2+e_B^2}}\int _0 ^{\infty}\mathrm{d}r.
	\end{split}
\end{equation}

We now consider perturbing the $x_1$ coordinate of the embedding as: $x_1= x _{\mathrm{cl}}(r) + \delta x_1(t,r)$, since every other perturbation is irrelevant to the stability. The Sturm-Liouville equation for the longitudinal perturbation $\delta x_1$ reads:

\begin{equation}\label{ddx1}
	- \frac{\mathrm{d}}{\mathrm{d}r}\left[\frac{(r^2-r_0^2)^{3/2}}{r} \frac{\mathrm{d}}{\mathrm{d}r} \right] \delta x_1(r) = \omega ^2 \left( \frac{8}{e_A^2+e_B^2} \right) ^2 \sqrt{1- \frac{r_0^2}{r^2}} \delta x_1(r) ,
\end{equation}
where we note that this is independent of $r_0$, as can be seen by changing variables to the dimensionless $\rho:=r/r_0$ as:

\begin{equation}\label{dxI}
	- \frac{\mathrm{d}}{\mathrm{d}\rho}\left[ \frac{(\rho^2-1)^{3/2}}{\rho} \frac{\mathrm{d}}{\mathrm{d}\rho}  \right]  \delta x_1(\rho) = \omega ^2 \left( \frac{8}{e_A^2+e_B^2} \right) ^2 \sqrt{1- \frac{1}{\rho^2}} \delta x_1(\rho) .
\end{equation}

\begin{figure}[H]
\centering
	\subfloat[$V ^{(\mathrm{I})} (\rho)$]{
		\includegraphics[width=65mm]{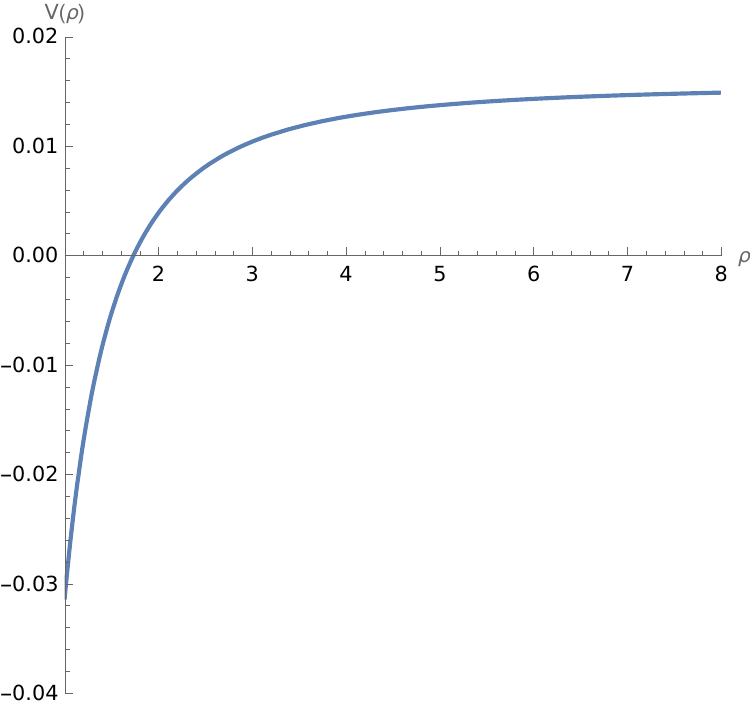}
}
	\subfloat[$V^{(\mathrm{I})} (y)$]{
		\includegraphics[width=65mm]{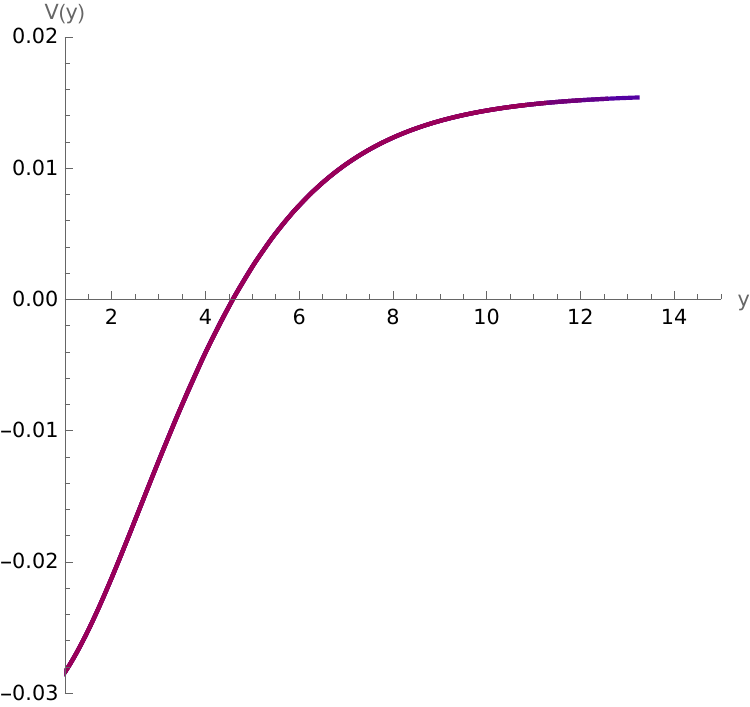}
}
	\caption{Schr\"{o}dinger potential of $\delta x_1$ for background $\mathrm{I}$ as a function of $\rho$ and $y(r)$ ($e_A=e_B=1$). We see that the potentials start from a negative value and after $\rho \approx 1.73$ become positive, being independent of $r_0$.}
	\label{VIdx}
\end{figure}

The variable to map \eqref{ddx1} to a Schr\"{o}dinger equation can be read from \eqref{cov} and in this case is:

\begin{equation}\label{yI}
	\begin{split}
		y(r;r_0) &= \frac{8}{e_A^2+e_B^2}\int _{r_0}^{r} \mathrm{d}\xi	\sqrt{\xi} \left( 1- \frac{r_0^2}{\xi^2} \right) ^{1/4} \left( \xi^2-r_0^2 \right) ^{-3/4}\\
		&=\frac{8}{e_A^2+e_B^2}\mathrm{arcoth}\left(  \frac{r}{\sqrt{r^2-r_0^2}}  \right) =\frac{8}{e_A^2+e_B^2}\mathrm{arcoth}\left(  \frac{\rho}{\sqrt{\rho^2-1}}  \right) ,
	\end{split}
\end{equation}

which defines the problem on the non-compact interval $[0,\infty)$ for $r\in [r_0,\infty)$ and is independent of $r_0$. The wavefunction of the Schr\"{o}dinger problem is related to $\delta x_1$ from \eqref{Psi} as:

\begin{equation}
	\Psi (y) = \sqrt{\frac{8}{ e_A^2+e_B^2}}\sqrt{\frac{\rho^2-1}{\rho}}\sqrt{r_0}\delta x_1(\rho) ,
\end{equation}

which vanishes for $r\to r_0$ ($\rho\to 1$) due to the prefactor, and for $r\to \infty$ satisfies $\Psi (y)\sim \sqrt{r}\delta x_1(r)$. The potential \eqref{V(r)} for this problem as a function of $r$ can be computed from \eqref{dxI} and the result is:

\begin{equation}
	V ^{(\mathrm{I})} (\rho) = \left( \frac{e_A^2+e_B^2}{8}\right)^2 \frac{\rho ^2-3}{4\rho^2},
\end{equation}
which asymptotes to $(e_A^2+e_B^2)^2/256=\pi^2/(4 L^{(\mathrm{I})}_{\mathrm{LST}})$ for large values of $\rho$ and is independent of $r_0$ as well. We can also look at the parametric graph of $V ^{(\mathrm{I})}(y)$, using $\rho = \rho (y)$ from \eqref{yI}. Both graphs of the potential as a function of the radial and the Schr\"{o}dinger variable are presented in Figure \ref{VIdx}.

\begin{figure}[H]
	\centering
	\includegraphics[width=0.8\textwidth]{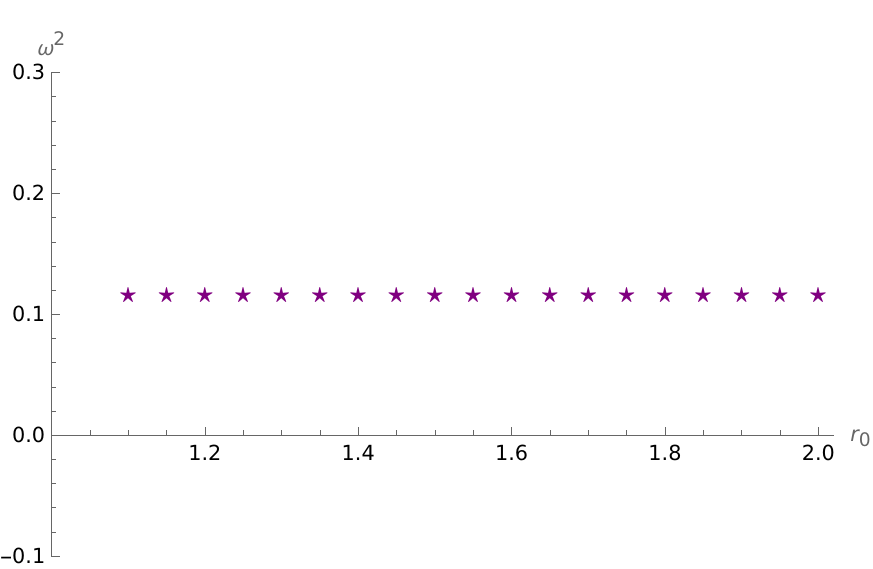}
	\caption{Plot of the lowest eigenvalues of the Schr\"{o}dinger problem for background I as a function of $r_0$, using the shooting method. We can see the independence of $r_0$ as the eigenvalue is constant, and also positive.}
	\label{Eigen1}
\end{figure}

The fact that \eqref{LI} was found to be a constant characteristic length, independent of the point on the tip of the string entering the bulk, is reflected in the equations of motion and the potentials of the longitudinal perturbations being also $r_0$-independent. Since there does not seem to be another configuration the system can change into, we expect no instabilities to exist for this background. In order to check the analysis, we also perform a numerical study of the eigenvalues for \eqref{ddx1} using the shooting method: We solve the equation separately in the intervals $[r _{\mathrm{IR}},r _{\mathrm{M}}]$ and $[r _{\mathrm{M}},r _{\mathrm{UV}}]$ where we took $r _{\mathrm{IR}}$ to be very close to $r_0$ and the middle point $r _{\mathrm{M}}$ to also be near $r _{\mathrm{UV}}$. The boundary conditions imposed were that the solution vanishes at $r=r _{\mathrm{UV}}$ and that \eqref{BCdx1} is satisfied at $r _{\mathrm{IR}}$, while the two solutions are then $"$matched$"$ with respect to the eigenvalue, in order to get the full solution on $[r _{\mathrm{IR}},r _{\mathrm{UV}}]$.
The results in Figure \ref{Eigen1} show the lowest eigenvalue $\omega^2 = \omega ^2 (r_0)$, the first one of which is positive, therefore no instability occurs.

Moving on to background II, the functions of the metric components are:

\begin{equation}
	\begin{split}
		&f ^{(\mathrm{II})} _x (r)= r^2,\\
		&f _{\varphi}^{(\mathrm{II})} (r)= r^2 f_s(r)+ 2\left(Q_A^2+Q_B^2\right)r^2\zeta^2(r),\\
		&g ^{(\mathrm{II})} (r)= \frac{4}{f_s(r)}\,\, , \,\, h^{(\mathrm{II})} (r)=g ^{(\mathrm{II})}(r)  ,\\
		&	F ^{(\mathrm{II})} (r)= \frac{4r^2}{f_s(r) \left( r^2-r_0^2 \right) }.
\end{split}
\end{equation}

Having again the same parametrization for the embedding, the classical solution reads:

\begin{equation}
	x^{(\mathrm{II})\prime} _{\mathrm{cl}} (r)= \pm \frac{2r_0}{r} \sqrt{\frac{1}{f_s(r)(r^2-r_0^2)}} = \pm \frac{2 \sqrt{2}r r_0}{\sqrt{(e_A^2+e_B^2)(r^2-r_0^2)(r^2-r _{+}^2)(r^2-r _{-}^2)}},
\end{equation}

and the length of the separation and energy are:

\begin{equation}\label{LII}
	\begin{split}
		L ^{(\mathrm{II})}(r_0) &= 4 r_0 \int _{r_0}^{\infty}\frac{\mathrm{d}r}{r\sqrt{f_s(r)(r^2-r_0^2)}}=\frac{4 \sqrt{2}r_0}{\sqrt{e_A^2+e_B^2}} \int _{r_0}^{\infty}\frac{\mathrm{d}r r}{\sqrt{(r^2-r_0^2)(r^2-r _{+}^2)(r^2-r _{-}^2)}}\\
		&= \sqrt{\frac{8}{e_A^2+e_B^2}}\frac{2}{\sqrt{1-\left( \frac{r _{-}}{r_0} \right) ^2}} \textbf{K}\left( \frac{r _{+}^2- r _{-}^2}{r_0^2 - r _{-}^2} \right)
	\end{split}
\end{equation}

\begin{equation}
	E ^{(\mathrm{II})}(r_0) = \frac{2}{\pi}\int _{r_0}^{\infty}\frac{ \mathrm{d}rr}{\sqrt{f_s(r)(r^2-r_0^2)}} - \frac{2}{\pi}\int _{r _{+}}^{\infty}\frac{\mathrm{d}r}{\sqrt{f_s(r)}},
\end{equation}

where $\textbf{K}(x)$ is the complete elliptic integral of the first kind\footnote{This is defined using the incomplete elliptic integral of the first kind as: $$\textbf{K}(x):=\textbf{F}\left( \left. \frac{\pi}{2}\right| x \right)\,\, ,\,\, \textbf{F}(\phi | x) = \int _0 ^{\phi}\frac{\mathrm{d}\theta}{\sqrt{1-x^2\sin^2\theta}}.$$}. Notice how the insertion of the function $f_s(r)$ in the square root of the denominator of \eqref{LII} forces the length to be non constant and have $r_0$ dependence. Also, for $r_0$ close to $r _{+}$ the length diverges as $(r-r _{+}) ^{-1}$ which means that string configurations reaching the end of the geometry give off infinite separation between the quark and anti-quark. This is yet another signal for confinement. As for the perturbations along the longitudinal direction, they obey the following equations of motion:

\begin{equation}
	- \frac{\mathrm{d}}{\mathrm{d}r} \left[ \frac{(r^2-r_0^2) ^{3/2} \sqrt{f_s(r)}}{2r} \frac{\mathrm{d}}{\mathrm{d}r}  \right] \delta x ^{(\mathrm{II})}(r) = \omega^2 2  \frac{\sqrt{r^2-r_0^2}}{r\sqrt{f_s(r)}}\delta x ^{(\mathrm{II})}(r)  .
\end{equation}

Once again, we formulate the Schr\"{o}dinger problem for $\delta x_1 ^{(\mathrm{II})}$, for which the Schr\"{o}dinger potential in terms of $r$ reads:

\begin{equation}\label{VIIx}
	\begin{split}
		V ^{(\mathrm{II})}(r;r_0)&=\frac{(r^2-3r_0^2)f_s(r) + r(r^2+r_0^2)f_s ^{\prime}(r)}{16r^2}\\
		&=\frac{e_A^2+e_B^2}{32r^6}\left\{ r^6-7r_0^2r _{-}^2r _{+}^2 + r^4(r _{+}^2 + r _{-}^2 - 3r_0^2) + r^2 \left[5r_0^2(r _{+}^2 + r _{-}^2) - 3r _{+}^2 r _{-}^2  \right]\right\},
	\end{split}
\end{equation}
and its asymptotic values are:

\begin{equation}
	\lim _{r\to \infty}V ^{(\mathrm{II})}(r;r_0) = \frac{e_A^2+e_B^2}{32},
\end{equation}

\begin{equation}
	\begin{split}
		\lim _{r_0\to r _{+}}\lim _{r\to r_0}V ^{(\mathrm{II})}(r;r_0) =\frac{e_A^2+e_B^2}{32r_{+}^6}&\Big[r_{+}^6-7r _{+}^4r _{-}^2+r_{+}^4\left(r _{-}^2-2r _{+}^2\right)\\
		&+r_{+}^2 \left( 5r_{+}^2(r _{+}^2+r _{-}^2) -3r _{+}^2r _{-}^2 \right)  \Big] 
	\end{split}
\end{equation}

where we make note that $V(r_0;r_0)$ approaches the negative value $-(e_A^2+e_B^2)/16$ as we increase $r_0$ and the behaviour of the potential is controlled by the parameters $e_A,e_B,r _{-}$ and $r _{+}$. One can also notice that in the case where $r _{+}^2=r _{-}^2$, for which four supercharges are preserved, $V(r_0;r_0)$ becomes zero\footnote{Interestingly, the choice $r _{+}^2=-r _{-}^2$ gives off $V(r_0;r_0)=(e_A^2+e_B^2)/4$.}.

\begin{figure}[H]
	\centering
	\subfloat[$r _{0}=1.2$]{
		\includegraphics[width=65mm]{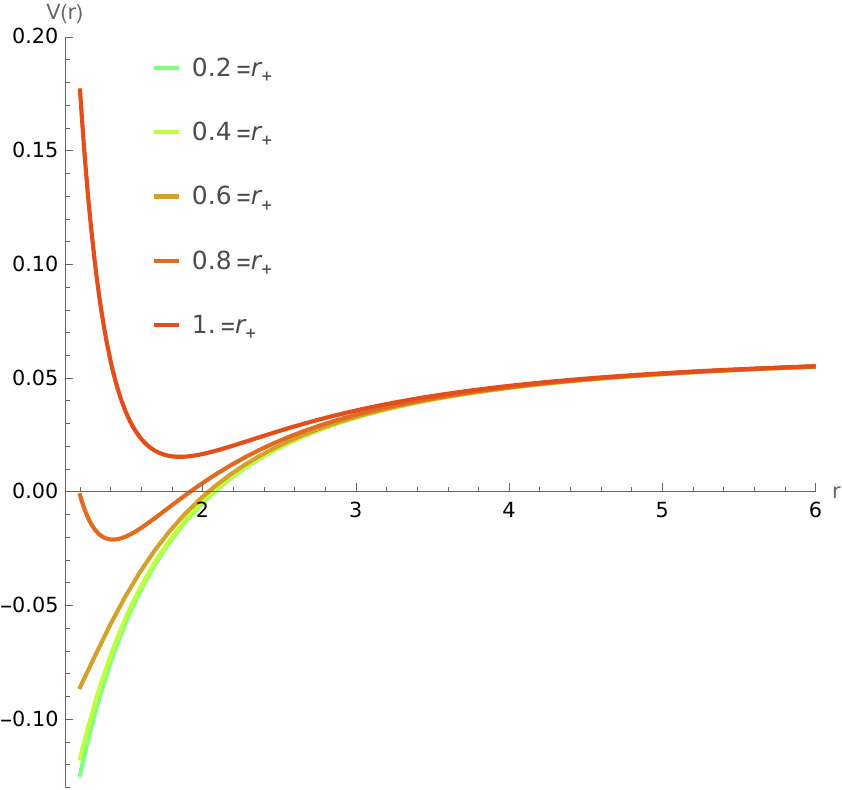}
		}
		\subfloat[$r _{0}=1.8$]{
			\includegraphics[width=65mm]{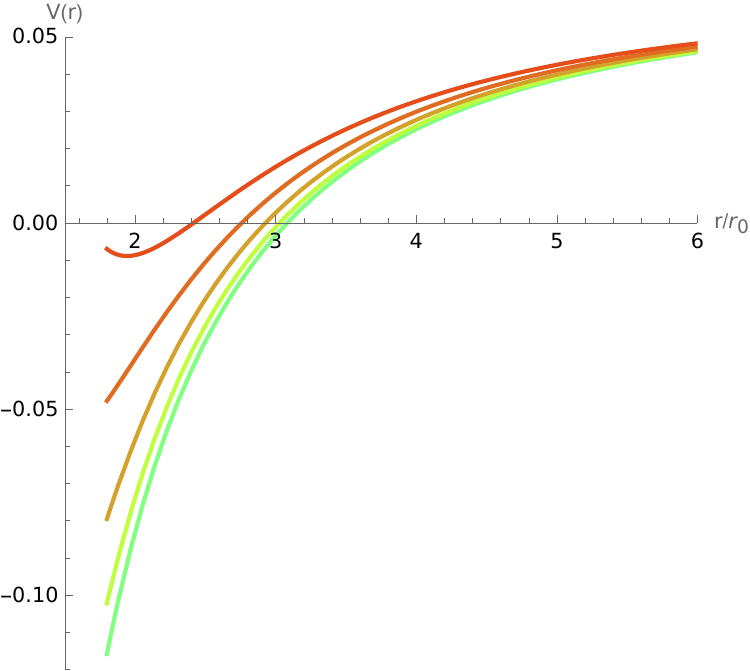}
			}

			\caption{Schr\"{o}dinger potential for the perturbations $\delta x$ in background II, for two values of $r _{0}$ ($e_A=e_B=1,r _{-}=0.2$) as $r _{+}$ also varies.}
			\label{VIIxplot}
\end{figure}

While the potential starts off with negative values for small $r$, which grow as $r_0$ gets larger, the numerical analysis shows that there are no negative eigenvalues.

\begin{figure}[H]
	\centering
	\includegraphics[width=0.73\textwidth]{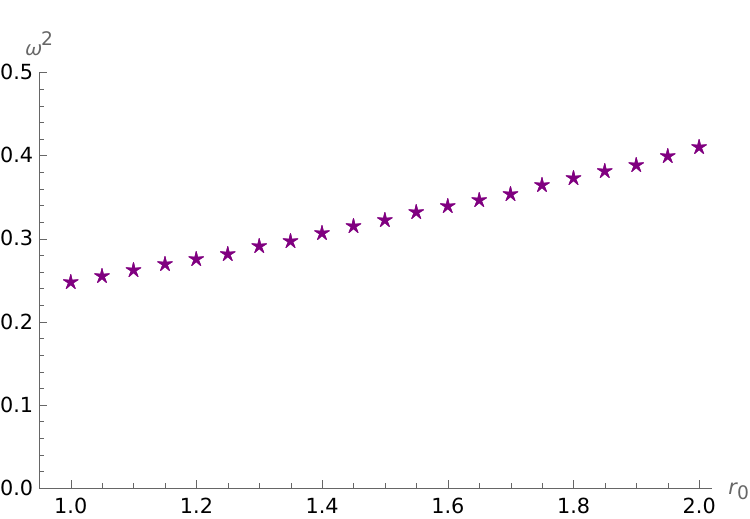}
	\caption{Lowest eigenvalues as a function of $r_0$ for the potential \eqref{VIIx} of background II (with $e_A=1=e_B$, $r _{-}=0.3,r _{+}=0.8$).}
	\label{VIIxeigen}
\end{figure}

\begin{figure}[H]
	\centering
	\includegraphics[width=0.74\textwidth]{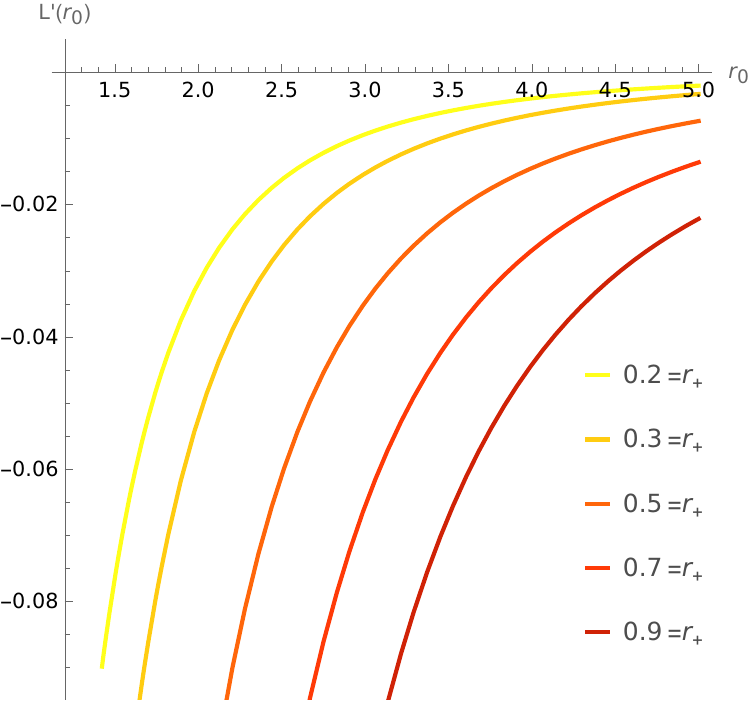}
	\caption{Plot of the derivative of the length function with respect to the turning point $r_0$, for various values of $r _{+}$ ($e_A=e_B=1,r _{-}=0.2$).}
	\label{L'II}
\end{figure}

We can also plot $L^{\prime}(r_0)$ (Figure \ref{L'II}) from \eqref{LII}, in order to see that it is manifestly negative and the concavity condition is satisfied.

There is also an approximate expression for the length of the separation, introduced in \cite{CMR1} which has the following form:

\begin{equation}\label{Lhat}
	\hat{L}(r_0)= \pi \left.\sqrt{g(r)} \left[ \frac{\mathrm{d}}{\mathrm{d}r} \left( \sqrt{f_x(r)} \right)  \right]^{-1} \right| _{r\to r_0}.
\end{equation}

For background ($\mathrm{II}$) we have:
\begin{equation}
	\hat{L} ^{(\mathrm{II})}(r_0) = \frac{2\pi}{\sqrt{f_s(r_0)}}= \sqrt{\frac{8}{e_A^2+e_B^2}} \frac{\pi r_0^2}{\sqrt{(r_0^2-r_{+}^2)(r_0^2-r _{-}^2)}},
\end{equation}

where we immediately see that for $r_0\to \infty$ we get the expression \eqref{LI} from background $\mathrm{I}$, indicating that the dual field theory of this background is also UV-completed by Little String Theory with this characteristic length, but due to the shrinking effect of the $\varphi$ circle, this theory confines. We also comment that this behaviour of the length asymptoting to a constant nonzero value is similar to the Maldacena-Nu$\tilde{\mathrm{n}}$ez \cite{MN} solution, which also is UV completed by the theory dual to $\mathrm{NS}5$ branes. This phenomenon acts as an indicator for non-local completions in general: A field theoretic UV completion would not be able to prevent the quark-anti-quark distance from becoming arbitrarily small as there is no characteristic scale, contrary to a theory of strings.    

\begin{figure}[H]
	\centering
	\subfloat[$r _{-}=0.2$]{
		\includegraphics[width=65mm]{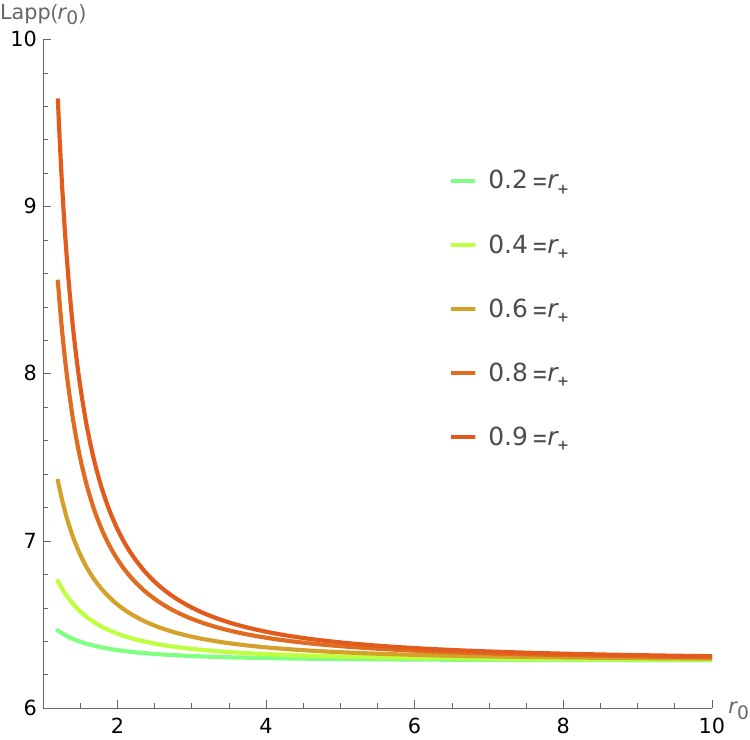}
		}
		\subfloat[$r _{-}=0.4$]{
			\includegraphics[width=65mm]{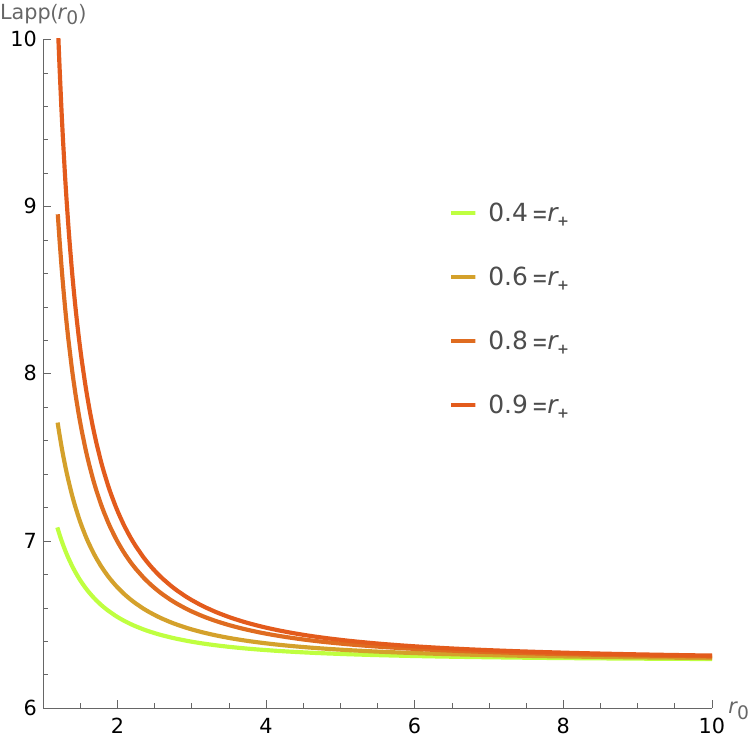}
			}
			\caption{Approximate form of the length for background II ($e_A=e_B=1$) for various values of $r_+$ and $r_{-}$.}
			\label{LhatII-varyingr0}
\end{figure}

Finally, we note that for the case where $m=0$ and $e_AQ_B=\pm e_BQ_A$, that is when the background preserves four supercharges, the potentials look very similar to Figure \ref{VIIxplot} and the numerical method also found the eigenvalues to be positive.

\subsubsection{'t Hooft loop}\label{thooft1}

We will now study the 't Hooft loop in background $\mathrm{II}$ by embedding a $\mathrm{D}5$ brane along\footnote{as before, we are using the $r$-gauge: $x=x(r)$ and $r=\sigma$.} $[t,r,\varphi,\theta _A, \phi_A,\psi _A]$ (we take $\hat{\mathrm{D}}5$ to wrap $\hat{\mathrm{S}}^3$ although the same is true for $\tilde{\mathrm{S}}^3$). Effectively, as the $\varphi$ cycle vanishes smoothly the object becomes $(1+1)$-dimensional, that is a string. The induced metric for the 't Hooft loop is:

\begin{equation}
	\begin{split}
		\mathrm{d}\mathrm{s}^2 _{\mathrm{ind}} = r &\left\{ - \mathrm{d}t^2 + \left[x ^{\prime 2}(r)  + \frac{4}{r^2 f _{s}(r)} \right]\mathrm{d}r^2  \right. \\
		& + \frac{2}{e_A^2}\left[ \hat{\omega} _1^2 + \hat{\omega} _2^2 + \left( \hat{\omega}_3 - e_A Q_A \zeta (r) \mathrm{d}\varphi \right) ^2  \right]\\
		& + \left[ f _{s}(r) + 2 Q_B^2 \zeta ^2 (r)  \right] \mathrm{d}\varphi ^2 \bigg\}, 
	\end{split}
\end{equation}
and the Born-Dirac-Infeld action can be written in the form:
\begin{equation}
	S _{\hat{\mathrm{D}}5} \left[x(r)  \right] = \int \mathrm{d}^6\sigma e ^{-\Phi} \sqrt{-\mathrm{det}(g _{\mathrm{ind}})}
\end{equation}
\begin{equation}
	\begin{split}
		&= \int _{\hat{\mathrm{S}}^3} \mathrm{d}\Omega _3 \int \mathrm{d}t \mathrm{d}r \mathrm{d}\varphi r^2 \sin \theta_A \sqrt{\left( x ^{\prime 2}(r) + \frac{4}{r^2 f_s(r)} \right) \left( f_s(r)+2 Q_B^2\zeta^2(r) \right) }\\
		&=\left( \frac{8}{e_A^2} \right) ^{3/2} L _{\varphi} T (4\pi)^2 \int _{r_0}^{\infty}\mathrm{d}r \sqrt{f _{x}(r) x^{\prime 2}(r) + g(r)},
	\end{split}
\end{equation}

where\footnote{$T=\int \mathrm{d}t$.} in this case, the functions read:

\begin{equation}
	\begin{split}
		&f^{(\mathrm{II})} _{t}(r) = r^4 \left[ f_s(r) + 2Q_B^2\zeta^2(r)  \right]\\
		&g^{(\mathrm{II})}_t(r) = 4r^2 \left[1 + \frac{2Q_B^2\zeta^2(r)}{f_s(r)}  \right] 
	\end{split}
\end{equation}

and we also take: 

\begin{equation}
	h ^{(\mathrm{II})}_t(r)=g^{(\mathrm{II})}_t(r)=4r^2 \left[1 + \frac{2Q_B^2\zeta^2(r)}{f_s(r)}  \right] 
\end{equation}

The classical solution then is

\begin{equation}
	\begin{split}
		x ^{\prime} _{\mathrm{cl}}(r) &= \pm \frac{2r_0^2}{r}\sqrt{\frac{f_s(r_0)+2Q_B^2\zeta^2(r_0)}{f_s(r)\left[r^4f_s(r)-r_0^4f_s(r_0)+2Q_B^2\left( r^4\zeta^2(r)-r_0^4\zeta^2(r_0) \right)   \right] }}\\
	\end{split}
\end{equation}

where $r_0$ is the turning point and $F$ is defined as in \eqref{F}. The length of the separation of the monopole-anti-monopole pair and energy in this case, are:

\begin{equation}
	\begin{split}
		L ^{(\mathrm{II})}_{\mathrm{MM}}(r_0) =&4r_0^2 \sqrt{f_s(r_0)+2Q_B^2\zeta^2(r_0)}\times\\
		&\int _{r_0}^{\infty} \frac{\mathrm{d}r}{r \sqrt{f_s(r) \left[r^4f_s(r)-r_0^4f_s(r_0) + 2Q_B^2 \left( r^4\zeta^2(r) - r_0^4\zeta^2(r_0) \right)   \right] }},
	\end{split}
\end{equation}

\begin{equation}
	\begin{split}
		E^{(\mathrm{II})} _{\mathrm{MM}}(r_0)&= \frac{2}{\pi}\int _{r_0}^{\infty}\mathrm{d}r \frac{r^3 \left[ f_s(r)+2Q_B^2 \zeta^2(r)  \right] }{\sqrt{f_s(r) \left[ r^4 f_s(r) - r_0^4f_s(r_0) + 2Q_B^2 \left( r^4 \zeta^2(r)-r_0^4\zeta^2(r_0) \right)   \right] }}\\
		&	-\frac{2}{\pi}\int _{r _{+}} ^{\infty}\mathrm{d}r r \sqrt{1+ \frac{2Q_B^2\zeta^2(r)}{f_s(r)}}
	\end{split}
\end{equation}

One can then check that $L^{\prime}(r_0)>0$ for every $r_0$, thus the embedding is deemed unstable. We can then formulate the Sturm-Liouville problem for the longitudinal perturbations of the magnetic string:

\begin{equation}
	-\frac{\mathrm{d}}{\mathrm{d}r} \left[P ^{(\mathrm{II})} _{\mathrm{MM}}(r;r_0) \frac{\mathrm{d}}{\mathrm{d}r}  \right]  \delta x_1(r) = \omega^2 W^{(\mathrm{II})} _{\mathrm{MM}}(r;r_0)\delta x_1(r),
\end{equation}

where:

\begin{equation}
	\begin{split}
		&P^{(\mathrm{II})} _{\mathrm{MM}}(r;r_0) =\frac{\sqrt{f_s(r)}\left[r^4(f_s(r)+2Q_B^2\zeta^2(r)) - r_0^4(f_s(r_0)+2Q_B^2\zeta^2(r_0))  \right]^{3/2}}{2r^3\left[ f_s(r) + 2Q_B^2\zeta^2(r)  \right]}\\
		&W^{(\mathrm{II})} _{\mathrm{MM}}(r;r_0) = \frac{2 \sqrt{r^4(f_s(r)+2Q_B^2\zeta^2(r)) - r_0^4(f_s(r_0) + 2Q_B^2\zeta^2(r_0))}}{\sqrt{f_s(r)}r^3 \left[f_s(r) + 2Q_B^2\zeta^2(r)  \right] }.
	\end{split}
\end{equation}

Using \eqref{V(r)} we can calculate (omitting the analytic expression of $V(r)$ since it is too complicated to write here explicitly) and plot the Schr\"{o}dinger potential of the magnetic string for various values of $r_0$ and $r _{+}$:

\begin{figure}[H]
	\centering
	\subfloat[$r _{+}=0.5$]{	\includegraphics[width=65mm]{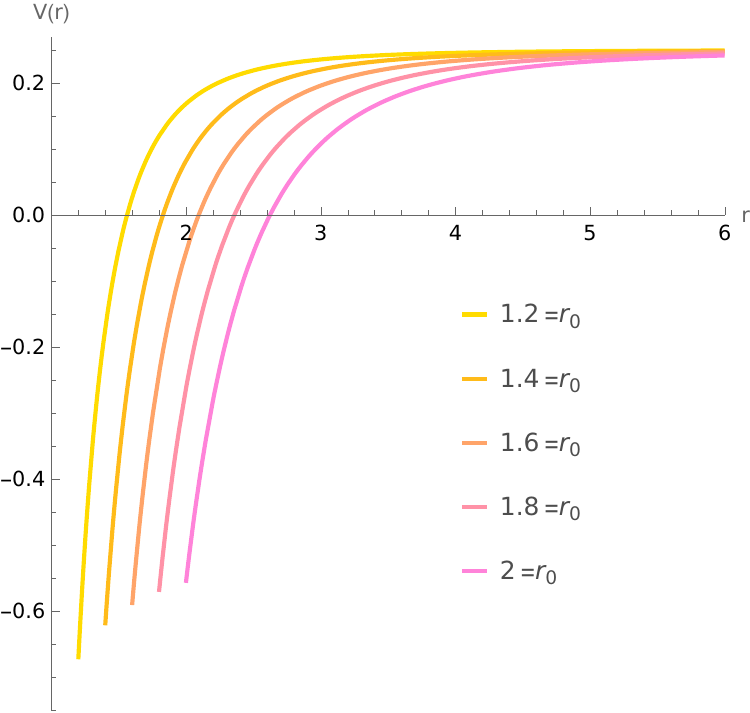}}
	\subfloat[$r _{0}=1.2$]{	\includegraphics[width=65mm]{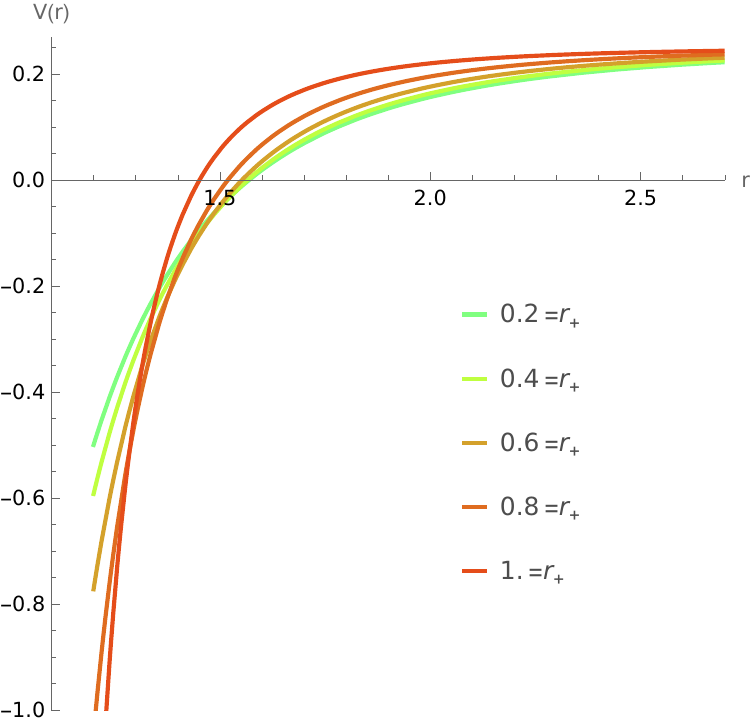}}
	\caption{Schr\"{o}dinger potential for the longitudinal perturbations of the 't Hooft loop, with various values of $r_0$ and $r _{+}$ (setting $r _{-}=0.2,r _{+}=0.5,Q_B=1=Q_A$).}
	\label{VtHooft}
\end{figure}

The potential asymptotes to $(e_A^2+e_B^2)/8$ for large values of the radial coordinate and as $r_0$ approaches the end of the space at $r _{+}$, it diverges to negative infinity for $r\to r_0$ for all values of the parameters. The later should indicate that there exist modes with negative eigenvalues, which means that the system is unstable under longitudinal fluctuations.\\

Studying the approximate length \eqref{Lhat} is also interesting, as the expression we get:

\begin{equation}
	\begin{split}
		\hat{L}^{(\mathrm{II})} _{\mathrm{MM}}(r_0) =& 2 \sqrt{2}\pi \frac{r _{+}^2 \sqrt{(r_0^2-r_+^2)\left[(e_A^2+e_B^2)(r_0^2-r _{-}^2)r _{+}^4 + 4Q_B^2 (r_0^2-r _{-}^2)  \right] }}{8Q_B^2 (r_0^2-r _{+}^2) -  (e_A^2+e_B^2)r _{+}^4(r _{+}^2+ r _{-}^2 - 2r_0^2)}\\
		&\times\sqrt{1 + \frac{4Q_B^2(r_0^2-r _{-}^2)}{(e_A^2+e_B^2)(r_0^2- r _{-}^2)r _{+}^4}},
	\end{split}
\end{equation}

satisfies:

\begin{equation}
	\lim _{r_0\to \infty}\hat{L}^{(\mathrm{II})} _{\mathrm{MM}}(r_0) = \pi \sqrt{\frac{2}{e_A^2+e_B^2}},
\end{equation}

\begin{equation}
	\hat{L}^{(\mathrm{II})} _{\mathrm{MM}}(r_0) \sim 8\pi \sqrt{\frac{r _{+}(r_0-r _{+})}{(e_A^2+e_B^2)(r _{+}^2 - r _{-}^2)}}\to 0 \,\, \,, \,\, \text{for} \,\, r_0\to r _{+}.
\end{equation}

We see that for the 't Hooft loop we get a maximum separation in the UV (since in this case the length which is monotonous increases from zero) again indicated by Little String Theory, that is half of the the minimum length we found for the Wilson loop. Also, as the string approaches the end of the space $r_{+}$, delving deeper into the bulk towards the far IR, the length goes to zero. This together with the fact that the effective tension for the dual string satisfies $T _{\mathrm{eff}}(r _{+})=r _{+}^2\sqrt{f_s(r _{+})+2Q_B^2\zeta^2(r _{+})}=0$ indicate screening of the monopole-anti-monopole pair: The system favors the disconnected configuration of the two magnetic strings extending infinitely into the bulk, whose endpoints at $r=\infty$ can move away from each other with no energy cost and therefore if one consideres the $U$ shaped embedding for the magnetic string, small fluctuations will destabilize the system.\\

\begin{figure}[H]
	\centering
	\subfloat[$r _{-}=0.2$]{	
	\includegraphics[width=60mm]{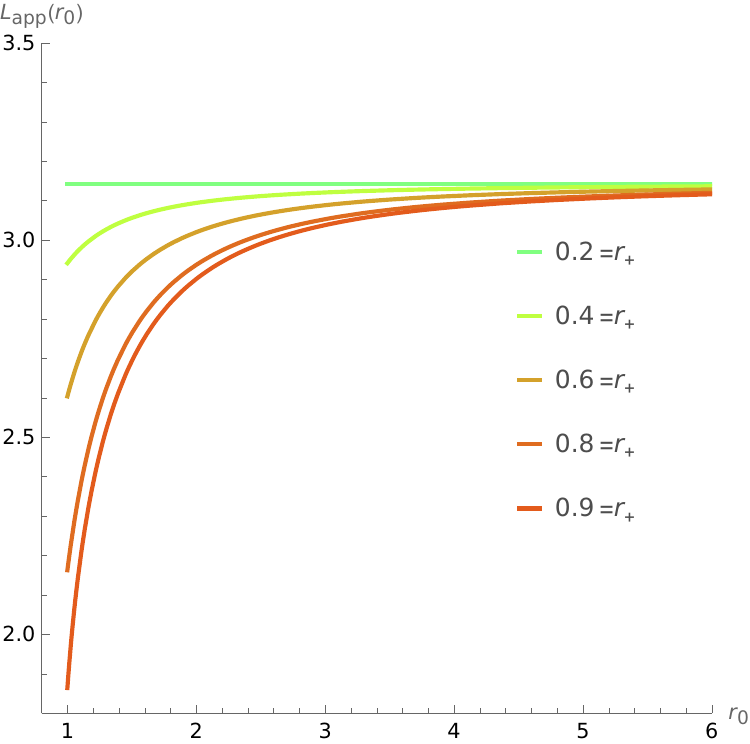}}
	\subfloat[$r _{-}=0.4$]{	\includegraphics[width=60mm]{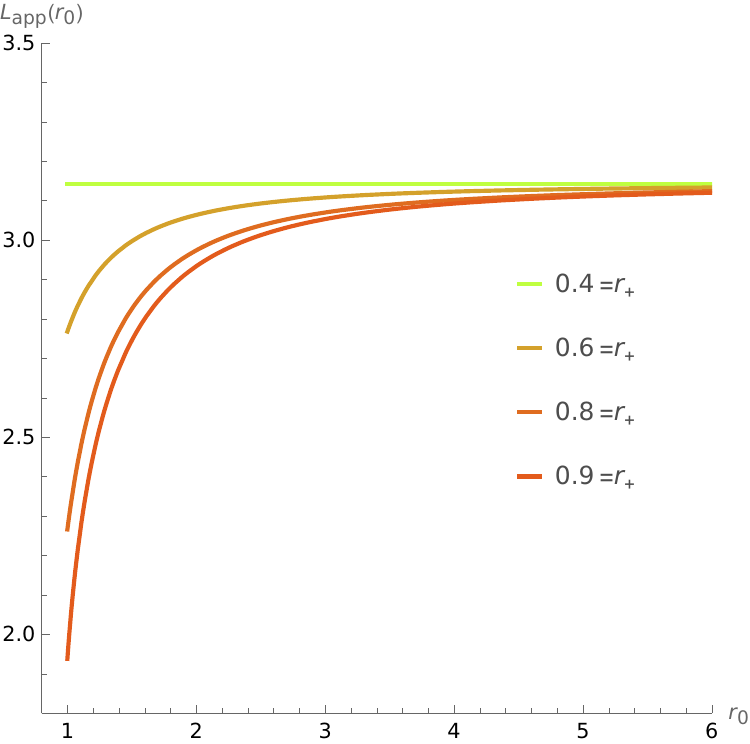}}
	\caption{Plot of the approximate length function for various $r _{+}$, whose first derivative is positive. It is apparent that the concavity condition does not hold, as the derivative is negative and then becomes zero.}
	\label{LapptII}
\end{figure}

To support the above argumets, we solve the equation numerically in order to get the $r_0$ dependence of the lowest eigenvalues using the shooting method, which indeed shows there are negative eigenvalues for small $r_0$:

\begin{figure}[H]
	\centering
	\subfloat[$r _{+}=0.36$]{	\includegraphics[width=65mm]{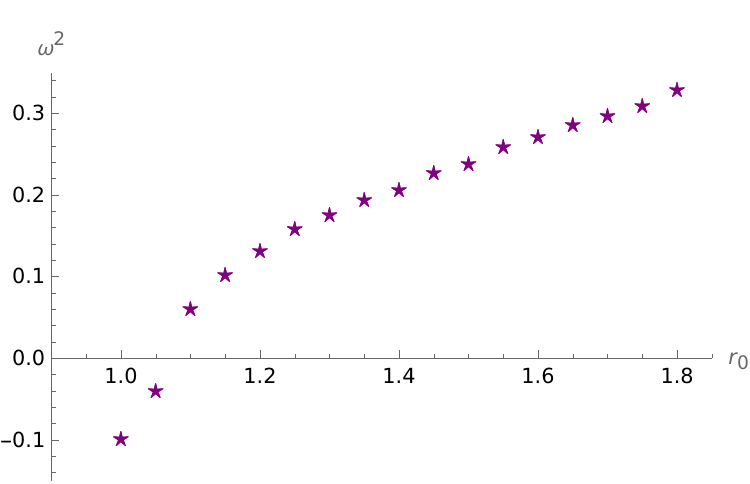}}
	\subfloat[$r _{+}=0.8$]{	\includegraphics[width=65mm]{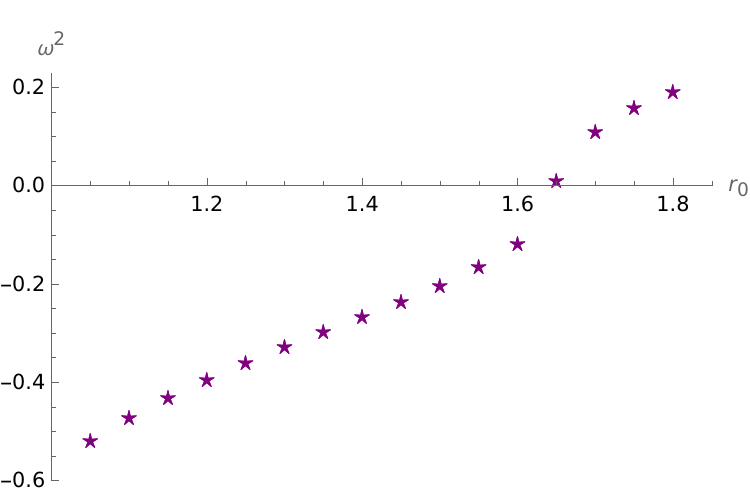}}
	\caption{Lowest eigenvalues for the 't Hooft loop as a function of $r_0$ for two values of $r _{+}$ (keeping $r _{-}=0.2$ and $e_A=e_B=1=Q_A=Q_B=1$).}
	\label{VtHooft}
\end{figure}

This behaviour is typical for backgrounds dual to confining theories, that is the Wilson loop is stable and confines while the 't Hooft loop is unstable and screens \cite{thooft1978}. This is in fact also the case for the Maldacena-Nu$\tilde{\mathrm{n}}$ez and Klebanov-Strassler solutions, which were studied using the same methods in \cite{Silva}. 

\subsubsection{Entanglement Entropy}
We now move to studying the perturbative stability of the Entangelment Entropy, for which we will choose the 8-manifold on which the metric \eqref{BGII} is induced to be spanned by the coordinates $[r,\varphi, \theta_A,\phi _A,\psi _A, \theta_B, \phi _B, \psi _B]$:

\begin{equation}
	\begin{split}
		\mathrm{d}\mathrm{s}^2 _{\Sigma _8} &= r \left\{ \left[ x ^{\prime 2}(r) + \frac{4}{r^2 f_s(r)}  \right] \mathrm{d}r^2 + f_s(r) \mathrm{d}\varphi^2   \right. \\
		&+ \frac{2}{e_A^2}\left[ \hat{\omega}_1^2+ \hat{\omega}_2^2 + \left( \hat{\omega}_3 - e_AQ_A \zeta (r)\mathrm{d}\varphi \right) ^2  \right]\\
		&\left.+ \frac{2}{e_B^2}\left[ \tilde{\omega}_1^2+ \tilde{\omega}_2^2 + \left( \tilde{\omega}_3 - e_BQ_B \zeta (r)\mathrm{d}\varphi \right) ^2  \right]\right\},
	\end{split}
\end{equation}

and the square root of the determinant of the induced metric is:

\begin{equation}
	\sqrt{e^{-4\Phi}\mathrm{det}(g _{\Sigma _8})}= \frac{8}{e_A^3e_B^3}\sqrt{r^4f_s(r) x ^{\prime 2}(r) + 4r^2}\sin\theta_A \sin\theta_B.
\end{equation}

Therefore the action is again reduced the the following integral:

\begin{equation}
	S ^{(\mathrm{II})} _{\mathrm{EE}} = \frac{2 (4\pi)^2 L _{\varphi}}{e_A^3e_B^3G_N} \int _{r_0} ^{\infty}\mathrm{d}r \sqrt{f^{(\mathrm{II})}_{\mathrm{EE}}(r) x^{(\mathrm{II})\prime 2}(r) + g^{(\mathrm{II})} _{\mathrm{EE}}(r)},
\end{equation}

where now,

\begin{equation}
	\begin{split}
		f^{(\mathrm{II})} _{\mathrm{EE}}(r) = r^4 f_s(r)\\
		g^{(\mathrm{II})} _{\mathrm{EE}}(r) = 4r^2=h^{(\mathrm{II})} _{\mathrm{EE}}(r).
	\end{split}
\end{equation}

The classical solution and the length read:

\begin{equation}
	x _{\mathrm{cl}} ^{(\mathrm{II})\prime}(r) = \pm 2 \frac{r_0^2}{r}\sqrt{\frac{f_s(r_0)}{f_s(r)\left[r^4f_s(r) - r_0^4f_s(r_0)  \right] }},
\end{equation}

\begin{equation}
	L^{(\mathrm{II})} _{\mathrm{EE}}(r_0) = 4r_0^2 \sqrt{f_s(r_0)}\int _{r_0}^{\infty} \frac{\mathrm{d}r}{r\sqrt{f_s(r)\left[r^4f_s(r)- r_0^4 f_s(r_0)  \right] }},
\end{equation}

where the derivative of the later is found to be manifestly positive, as in the 't Hooft loop, which is an indication of instability. The Sturm-Liouville equation for the longitudinal perturbation of this problem is then:

\begin{equation}
	-\frac{\mathrm{d}}{\mathrm{d}r} \left[P^{(\mathrm{II})} _{\mathrm{EE}}(r;r_0) \frac{\mathrm{d}}{\mathrm{d}r}  \right]  \delta x_1(r) = \omega^2 W^{(\mathrm{II})} _{\mathrm{EE}}(r;r_0)\delta x_1(r),
\end{equation}

where:

\begin{equation}
	\begin{split}
		&P^{(\mathrm{II})} _{\mathrm{EE}}(r;r_0) =\frac{\left[r^4f_s(r) - r_0^4f_s(r_0)  \right]^{3/2} }{2r^3 \sqrt{f_s(r)}}\\
		&W^{(\mathrm{II})} _{\mathrm{EE}}(r;r_0) =\frac{2 \sqrt{r^4f_s(r) - r_0^4f_s(r_0)}}{r \sqrt{f_s(r)}}.
	\end{split}
\end{equation}

Again, we can get the Schr\"{o}dinger potential from \eqref{V(r)} and plot it for various $r_0$ to find that it too expresses instability:

\begin{figure}[H]
	\centering
	\includegraphics[width=0.75\textwidth]{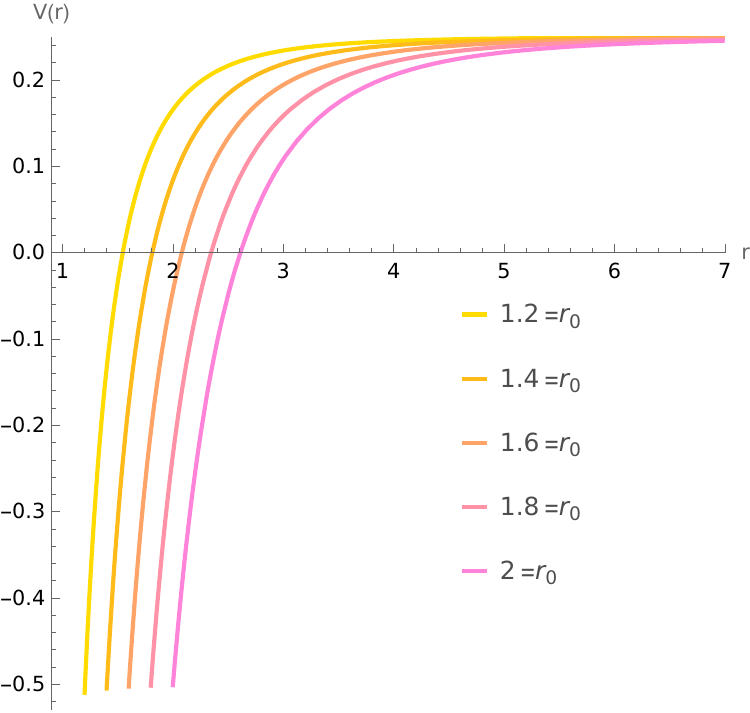}
	\caption{Schr\"{o}dinger potential for longitudinal perturbations of the Entanglement Entropy for various $r_0$ (fixing $r _{-}=0.2, r _{+}=0.5, e_A=1=e_B$).}
	\label{VEE}
\end{figure}
The approximate length reads:

\begin{equation}
	\hat{L}^{(\mathrm{II})} _{\mathrm{EE}}(r_0) = \pi \sqrt{\frac{8}{e_A^2+e_B^2}} \frac{\sqrt{(r_0^2-r _{-}^2)(r_0^2 - r _{+}^2)}}{2r_0^2 - r _{-}^2 - r _{+}^2}.
\end{equation}

This approaches the same maximum length with the 't Hooft loop as $r_0\to \infty$ , that is half of \eqref{LI} and $\hat{L} _{\mathrm{EE}}\to 0$ as $r_0\to r _{+}$. We thus have a similar behaviour as the monopole-anti-monopole scenario, namely the system screens and the preferable configuration is the disconnected one. We note that for the 't Hooft loop and Entanglement Entropy, due to the fact that the functions expressing the shrinking $\mathrm{S}^1[\varphi]$ and the fibration with $\mathrm{S}^3[\theta,\phi,\psi]$ both vanish at $r=r _{+}$: $f_s(r _{+})=0=\zeta (r _{+})$, the effective tension of the dual magnetic flux tube vanishes as $r_0\to r_{+}$, which agrees with our results.

\subsection{Background dual to (4+1)-dimensional QFTs}

We will now study the stability of our observables in background $\mathrm{III}$, which is dual to five-dimensional theory at low energies.

\subsubsection{Wilson loop}
The functions of the metric read:

\begin{equation}
	\begin{split}
		f ^{(\mathrm{III})}_{x_1}(r) = r^2, \,\, g^{(\mathrm{III})}(r) = \frac{N}{f_s(r)} = h^{(\mathrm{III})}(r), \\
		f^{(\mathrm{III})} _{\varphi}(r) = r^2\left[ f_s(r) + 2Q^2\zeta^2(r)  \right] ,
	\end{split}
\end{equation}

and the classical solution is:

\begin{equation}
	x^{(\mathrm{III})\prime} _{1\mathrm{cl}}(r) =\pm \frac{r_0}{r}\sqrt{\frac{N}{f_s(r)(r^2-r_0^2)}}=\pm r r_0 \sqrt{\frac{N}{(r^2-r_0^2)(r^2-r _{+}^2)(r^2-r _{-}^2)}},
\end{equation}

or, after integration:

\begin{equation}
	\begin{split}
		x^{(\mathrm{III})} _{1\mathrm{cl}}(r) = \mp & \frac{Nr_0}{(r _0^2 - r _{-}^2)(r_0^2- r _{+}^2)(r _{-}^2 - r _{+}^2)}\left[ r_0^3 (r _{-}^2- r _{+}^2) \arctanh \left( \frac{r}{r_0} \right)+\right.\\
		&\left. r _{-}^4 (r _{+}^2 - r _{0}^2)\arctanh \left( \frac{r }{r _{-}} \right) + r _{+}^3 (r _0^2 - r _{-}^2  ) \arctanh \left( \frac{r}{r _{+}} \right)\right] ,
	\end{split}
\end{equation}

which gives off the expression for the length:

\begin{equation}
	L^{(\mathrm{III})}(r_0) = 2 \sqrt{N}r_0 \int _{r_0} ^{\infty} \frac{\mathrm{d}r r}{\sqrt{(r^2-r _{+}^2)(r^2-r _{-}^2)(r^2-r_0^2)}}.
\end{equation}

The approximate expression \eqref{Lhat} for this case is:

\begin{equation}
	\hat{L} ^{(\mathrm{III})} (r_0) = \frac{\sqrt{N}\pi}{\sqrt{f_s(r_0)}}=\frac{\pi \sqrt{N}r_0^2}{\sqrt{(r_0^2-r _{+}^2)(r_0^2-r _{-}^2)}},
\end{equation}

which, up to a constant, is the same as the case of background $\mathrm{II}$, signifying that the Wilson loop in this background has the same behaviour: We get a configuration with minimum separation dictated by Little String Theory for $r_0\to\infty$ and a divergence for $r_0\to r _{+}$, that is stable ($L^{\prime}(r_0)<0$). The equation of motion for the longitudinal fluctuations in this case is:

\begin{equation}
	- \frac{\mathrm{d}}{\mathrm{d}r}\left[ P ^{(\mathrm{III})}(r;r_0) \frac{\mathrm{d}}{\mathrm{d}r}  \right] \delta x_1(r) = \omega^2 W ^{(\mathrm{III})}(r;r_0)\delta x_1 (r) ,
\end{equation}

where:

\begin{equation}
	\begin{split}
		& P ^{(\mathrm{III})}(r;r_0) = \frac{(r^2-r_0^2)^{3/2}\sqrt{(r^2-r _{-}^2)(r^2-r _{+}^2)}}{r^3 \sqrt{N}},\\
		& W ^{(\mathrm{III})}(r;r_0) = \frac{\sqrt{N}r\sqrt{r^2-r_0^2}}{\sqrt{(r^2-r _{-}^2)(r^2- r _{+}^2)}},
	\end{split}
\end{equation}

and the potential reads:

\begin{equation}
	V ^{(\mathrm{III})}(r;r_0)= \frac{r^6-7r_0^2r _{+}^2 r _{-}^2 + r^4 \left( r _{+}^2+ r _{-}^2 - 3r_0^2 \right) +r^2 \left[ 5r_0^2 \left(r _{+}^2 + r _{-}^2  \right) -3r _{+}^2 r _{-}^2\right]}{4 N r^6}.
\end{equation}

It is clear that for large radial coordinates this asymptotes to the value $\frac{1}{4N}$ which gets smaller as the number of branes grows, while for $r$ close to $r_0$ as $r_0$ approaches $r _{+}$, we get:

\begin{equation}
	\lim _{r_0\to r _{+}}V ^{(\mathrm{III})}(r_0;r_0)= \frac{r _{+}^6+ r _{+}^4(r _{-}^2-2r _{+}^2) + r _{+}^2 \left[5 r _{+}^2(r _{+}^2+r _{-}^2)-3r _{+}^2 r _{-}^2  \right] - 7r _{+}^4 r _{-}^2 }{4Nr _{+}^6}
\end{equation}

\begin{figure}[H]
	\centering
	\subfloat[$r _0=1.2$]{
		\includegraphics[width=65mm]{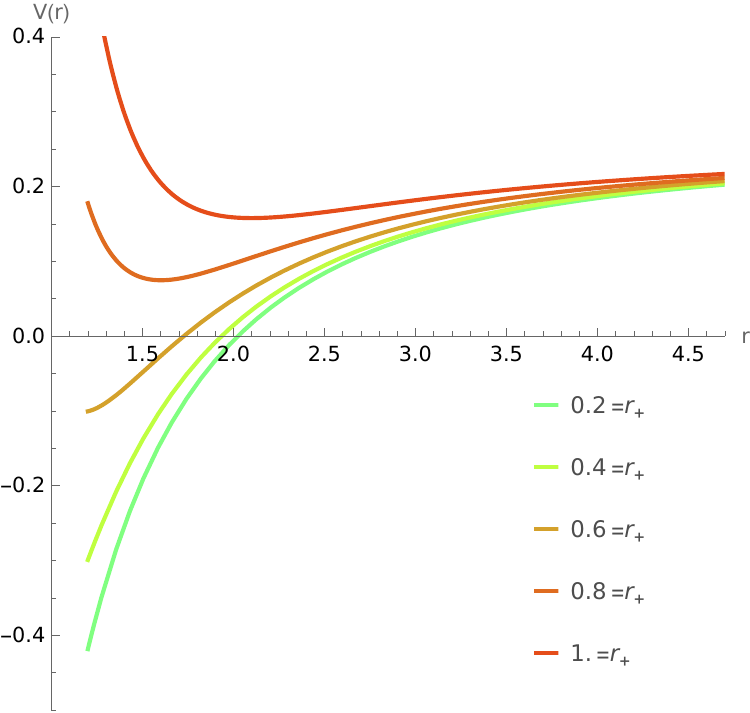}
		}
		\subfloat[$r _0=1.8$]{
			\includegraphics[width=65mm]{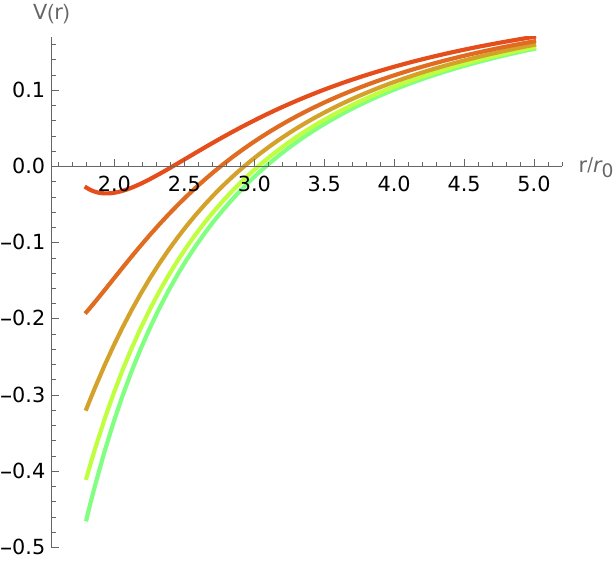}
			}
			\caption{Schr\"{o}dinger potential for longitudinal perturbations in background $\mathrm{III}$, for $r_0=1.2$ and $r_0=1.8$ as $r _{+}$ varies ($r_{-}=0.2$,$N=1$).}
			\label{V3x}
\end{figure}

\begin{figure}[H]
	\centering
	\includegraphics[width=0.8\textwidth]{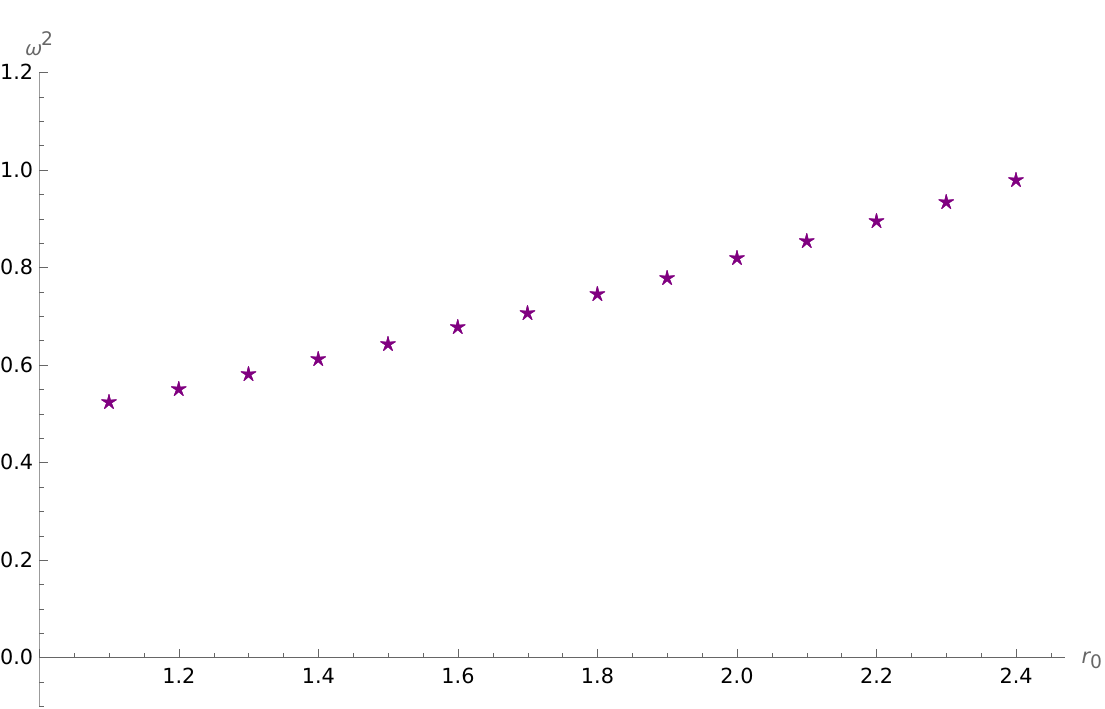}
	\caption{Lowest eigenvalues as a function of $r_0$ for the Wilson loop of background $\mathrm{III}$ ($r _{-}=0.2$, $r _{+}=0.2$, $N=1$).}
	\label{eee}
\end{figure}

Even though the potential explores a small negative region for larger values of $r_0$ (Figure \ref{V3x}) it is evident from the numerical results in Figure \ref{eee} that no negative eigenvalue modes exist.

\subsubsection{'t Hooft loop}

For the 't Hooft loop, we use the same procedure as in Section \ref{thooft1} and we find:

\begin{equation}
	S _{\hat{D}5}= L _{\varphi} (4\pi)^2 T \left( \frac{N}{4} \right) ^{3/2}\int _{r_0} ^{\infty}\mathrm{d}r \sqrt{f_x(r) x ^{\prime 2}(r) + g(r)},
\end{equation}

and the functions are:
\begin{equation}
	\begin{split}
		&	f^{(\mathrm{III})}_t (r) = r^4f_s(r) , \,\, g^{(\mathrm{III})}_t(r) = r^2N=h^{(\mathrm{III})}_t(r), \\
		&F^{(\mathrm{III})}_t(r) = \frac{r^6f_s(r)N}{r^4f_s(r)-r_0^4f_s(r_0)}.
	\end{split}
\end{equation}

The length of the monopole-anti-monopole separation  and its approximate expression are then found to be:

\begin{equation}\label{Lt3}
	L ^{(\mathrm{III})} _{\mathrm{MM}}(r_0) = 2 r_0^2\sqrt{Nf_s(r_0)}\int _{r_0} ^{\infty}\mathrm{d}r \frac{1}{r \sqrt{f_s(r)\left[r^4f_s(r) - r_0^4f_s(r_0)\right]}},
\end{equation}

\begin{equation}
	\hat{L} ^{(\mathrm{III})} _{\mathrm{MM}}(r_0) = \pi \sqrt{N} \frac{\sqrt{(r_0^2-r _{+}^2)(r_0^2-r _{-}^2)}}{2r_0^2 - r _{+}^2 - r _{-}^2}.
\end{equation}

\begin{figure}[H]
	\centering
	\subfloat[$r _{+}=0.3$]{
		\includegraphics[width=65mm]{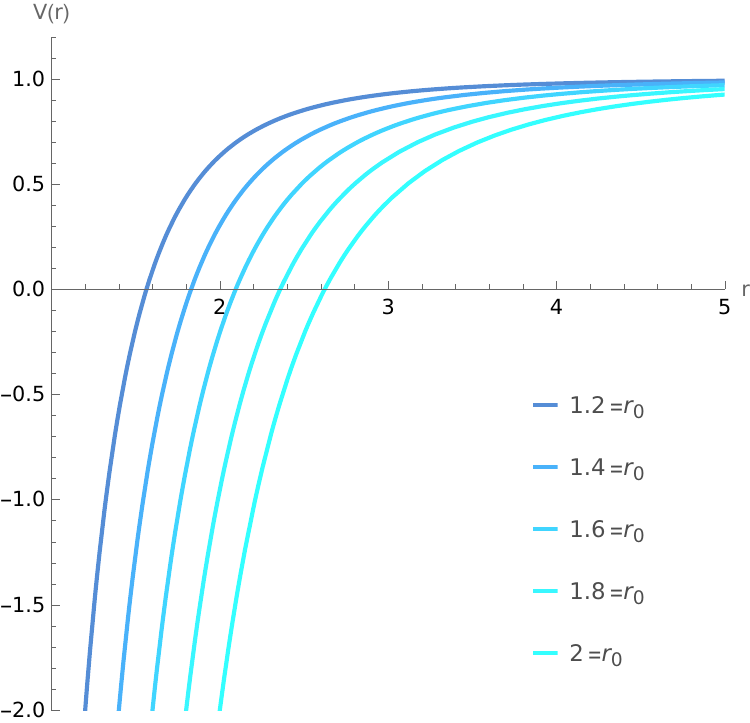}
		}
		\subfloat[$r _0=1.2$]{
			\includegraphics[width=65mm]{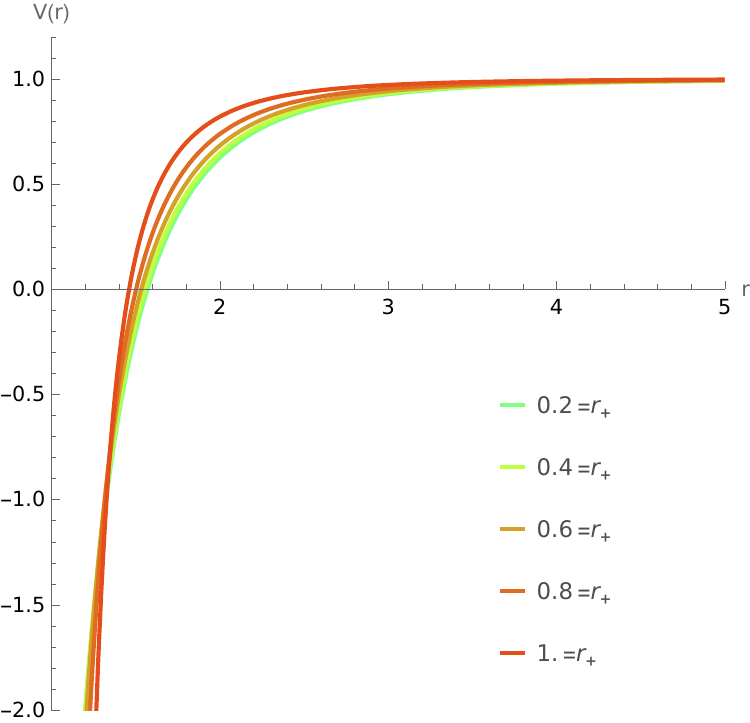}
			}
			\caption{Schr\"{o}dinger potential for longitudinal perturbations of the 't Hooft loop in background $\mathrm{III}$ as $r_0$  and $r _{+}$ vary, for fixed values of parameters ($N=1$ and $r _{-}=0.2$).}
			\label{V3tr0vary}
\end{figure}

From this we see that again, as in background II, we get a maximal length as $r_0\to\infty$ and a vanishing one in the IR as $r_0\to r _{+}$. Indeed, we find the plot in Figure \ref{V3tr0vary} to be of the same form as the one in background II, that is, there exist negative eigenvalues $\omega^2$ and the configuration is unstable: It is preferable for the system to move into the disconnected solution of the two $\mathrm{D}5$ branes which they can move away from each other freely, therefore we expect a screening scenario. \\

The perturbations obey:

\begin{equation}\label{tHooftIII}
	- \frac{\mathrm{d}}{\mathrm{d}r}\left[ P_{\mathrm{MM}} ^{(\mathrm{III})}(r;r_0) \frac{\mathrm{d}}{\mathrm{d}r}  \right] \delta x_1(r) = \omega^2 W_{\mathrm{MM}} ^{(\mathrm{III})}(r;r_0)\delta x_1 (r) ,
\end{equation}

whith:

\begin{equation}
	P_{\mathrm{MM}} ^{(\mathrm{III})}(r;r_0) = \frac{\left[(r^2-r_0^2)(r^2+r_0^2-r _{-}^2 - r _{+}^2)  \right]^{3/2} }{r \sqrt{N}\sqrt{(r^2- r _{+}^2)(r^2- r _{-}^2)}},
\end{equation}

\begin{equation}
	W_{\mathrm{MM}} ^{(\mathrm{III})}(r;r_0) =\frac{r \sqrt{N}\sqrt{(r^2-r_0^2)(r^2+r_0^2-r _{-}^2 - r _{+}^2)}}{\sqrt{(r^2-r _{+}^2)(r^2- r _{-}^2)}}.
\end{equation}

\begin{figure}[H]
	\centering
	\includegraphics[width=0.75\textwidth]{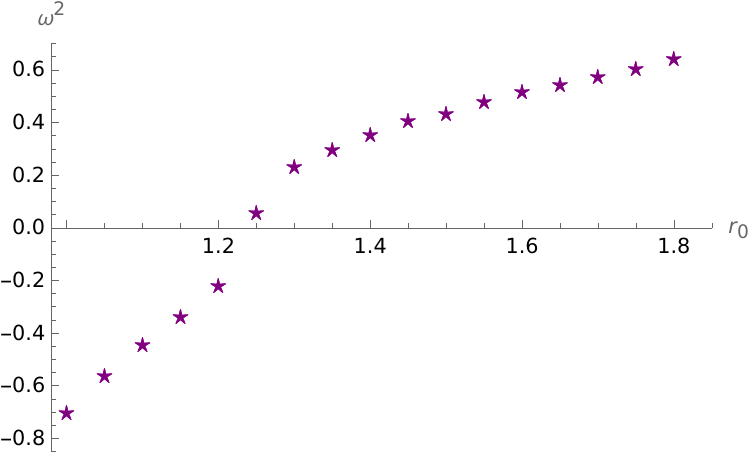}
	\caption{Plot of the lowest eigenvalues $\omega^2(r_0)$ of the Schr\"{o}dinger equation for background III ($r _{-}=0.2$, $r _{+}=0.8$ and $N=1$).}
	\label{Eigen4}
\end{figure}

The potentials depicted in Figure \ref{V3tr0vary} and the numerical solution in Figure \ref{Eigen4} confirm the claims about screening, as the system is unstable under small perturbations and transitions to the disconnected configuration.

\subsubsection{Entanglement Entropy}

The induced metric on the 8-manifold $\Sigma _{8}$ spanned by $\left[x_1,x_2,x_3,x_4,\varphi,\theta,\phi,\psi\right]$ after letting $x_1=x_1(r)$ is:

\begin{equation}
	\begin{split}
		\mathrm{d}s ^2 _{\Sigma _8} &= r \left\{\mathrm{d}x_2^2+\mathrm{d}x_3^2 + \mathrm{d}x_4 ^2 + \left( \frac{N}{r^2f_s(r)}+x _1 ^{\prime 2}\right)\mathrm{d}r^2\right.\\
		& \left.+\frac{N}{4}\left[ \omega_1^2 + \omega_2^2 + \left(  \omega_3 - \sqrt{\frac{8}{N}} Q\zeta(r) \mathrm{d}\varphi \right) ^2  \right] 	\right\} .
	\end{split}
\end{equation}

The action is then found to be:

\begin{equation}
	S _{\mathrm{EE}}^{(\mathrm{III})} = \frac{(4\pi)^2N^{3/2}}{8G_N}L _{\varphi}L _{x_2}L _{x_3}L _{x_4} \int _{r_0} ^{\infty}\mathrm{d}r \sqrt{f _{\mathrm{EE}} ^{(\mathrm{III})}(r)x ^{\prime 2}(r) + g _{\mathrm{EE}} ^{(\mathrm{III})}(r)},
\end{equation}

where

\begin{equation}
	f _{\mathrm{EE}} ^{(\mathrm{III})} = r^4 f_s(r) , \,\, g _{\mathrm{EE}} ^{(\mathrm{III})}= r^2N, \,\,F _{\mathrm{EE}}^{(\mathrm{III})}= \frac{r^6N f_s(r)}{r^4f_s(r) - r_0^4f_s(r_0)}
\end{equation}

The length in this case is given by the following integral, which we notice is the exact same expression with the 't Hooft loop \eqref{Lt3}:
\begin{equation}
	L _{\mathrm{EE}} ^{(\mathrm{III})} (r_0) = 2r_0^2 \sqrt{Nf_s(r_0)} \int _{r_0}^{\infty}\frac{\mathrm{d}r}{r \sqrt{f_s(r) \left[r^4f_s(r) - r_0^4f_s(r_0)  \right] }},
\end{equation}

so the same screening behaviour will take place, as the equations of motion (and thus the potential) of the fluctuations are also the same as in the 't Hooft loop case \eqref{tHooftIII}.

\subsection{Background dual to (2+1)-dimensional QFTs}
Lastly, we will study background $\mathrm{IV}$ which produces at low energies a $(2+1)$-dimensional effective theory.

\subsubsection{Wilson loop}
The functions from the metric of this geometry are:

\begin{equation}
	\begin{split}
		&f^{(\mathrm{IV})} _{x_1}(r)=r^2 \,\, ,\,\, g^{(\mathrm{IV})} (r)= \frac{2r^2}{  r^2 - m} = h ^{(\mathrm{IV})}(r)\\
		&f _{\varphi} ^{(\mathrm{IV})}(r) = -r^2 \sin ^2\vartheta \,\,, \,\, f _{\vartheta}^{(\mathrm{IV})}(r)= -r^2\,\,, \,\, f^{(\mathrm{IV})} _{\mu}(r)=(r^2-m^2),
	\end{split}
\end{equation}

and the classical solution:

\begin{equation}
	x ^{(\mathrm{IV})\prime} _{1\mathrm{cl}}(r)= \pm r _0 \sqrt{\frac{2}{(r^2-r_0^2)(r^2-m)}}.
\end{equation}

The length of the separation and energy of the quark-anti-quark pair are then given by the integrals:

\begin{equation}
	\begin{split}
		L ^{(\mathrm{IV})}(r_0) &= 2 \sqrt{2}r_0 \int _{r_0} ^{\infty} \frac{\mathrm{d}r}{\sqrt{(r^2-r_0^2)(r^2-m)}} \\
		&= 2 \sqrt{2}\textrm{\textbf{F}}\left( \arcsin \left( \frac{r_0}{\sqrt{m}}\right)\Big|\frac{m}{r_0^2} \right)+2i \sqrt{2}\mathrm{\textbf{K}}\left( 1- \frac{m}{r_0^2} \right)    ,
	\end{split}
\end{equation}

\begin{equation}
	E^{(\mathrm{IV})}(r_0)= \frac{\sqrt{2}}{\pi}\int _{r_0}^{\infty}\frac{\mathrm{d}r r^2}{\sqrt{(r^2-r_0^2)(r^2-m)}}- \frac{\sqrt{2}}{\pi}\int _m ^{\infty} \frac{\mathrm{d}r r}{\sqrt{r^2-m}},
\end{equation}

and the perturbations obey the following equation of motion:

\begin{equation}
	- \frac{\mathrm{d}}{\mathrm{d}r}\left[ P ^{(\mathrm{IV})}(r;r_0) \frac{\mathrm{d}}{\mathrm{d}r}  \right] \delta x_1(r) = \omega^2 W ^{(\mathrm{IV})}(r;r_0)\delta x_1 (r) ,
\end{equation}

where:

\begin{equation}
	\begin{split}
		& P ^{(\mathrm{IV})}(r;r_0) = \frac{\sqrt{r^2-m}(r^2-r_0^2)^{3/2}}{\sqrt{2}r^2},\\
		&W ^{(\mathrm{IV})}(r;r_0) = \sqrt{\frac{2(r^2-r_0^2)}{r^2-m}}.
	\end{split}
\end{equation}

The Schr\"{o}dinger potential for the above reads:

\begin{equation}
	V ^{(\mathrm{IV})}(r;r_0)= \frac{r^4+(m-3r_0^2)r^2+5mr_0^2}{8r^4},
\end{equation}

which has the limiting behaviour:

\begin{equation}
	\begin{split}
		&\lim _{r\to\infty} V ^{(\mathrm{IV})}(r;r_0)= \frac{1}{8},\\
		&\lim _{r_0\to \sqrt{m}}V ^{(\mathrm{IV})}(r_0;r_0)= \frac{1}{2}.
	\end{split}
\end{equation}

\begin{figure}[H]
	\centering
	\subfloat[$m=1$]{
		\includegraphics[width=65mm]{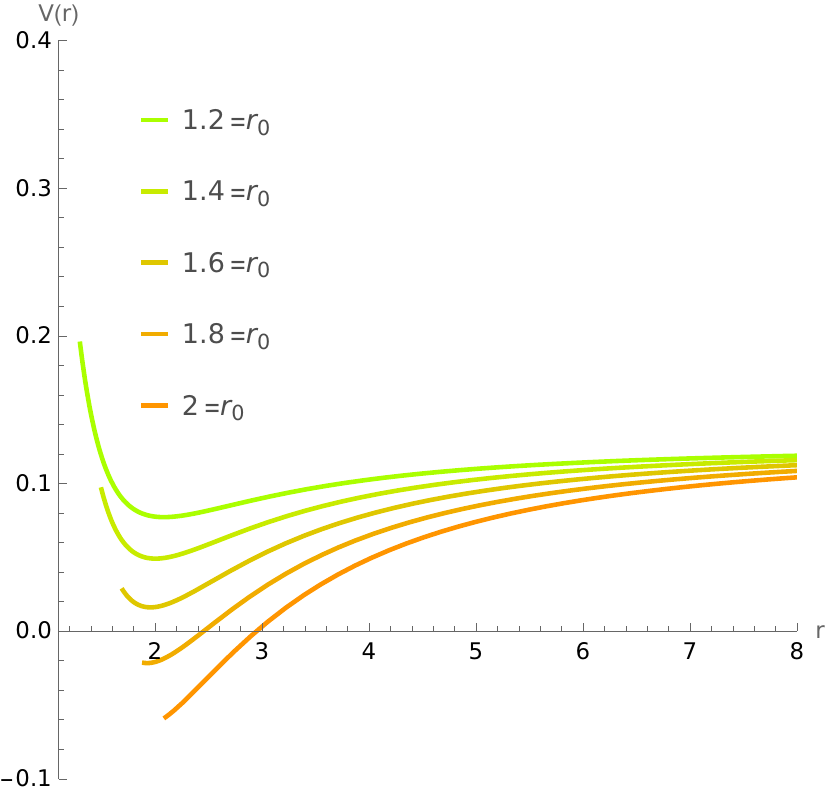}
		}
		\subfloat[$r _0=1.2$]{
			\includegraphics[width=65mm]{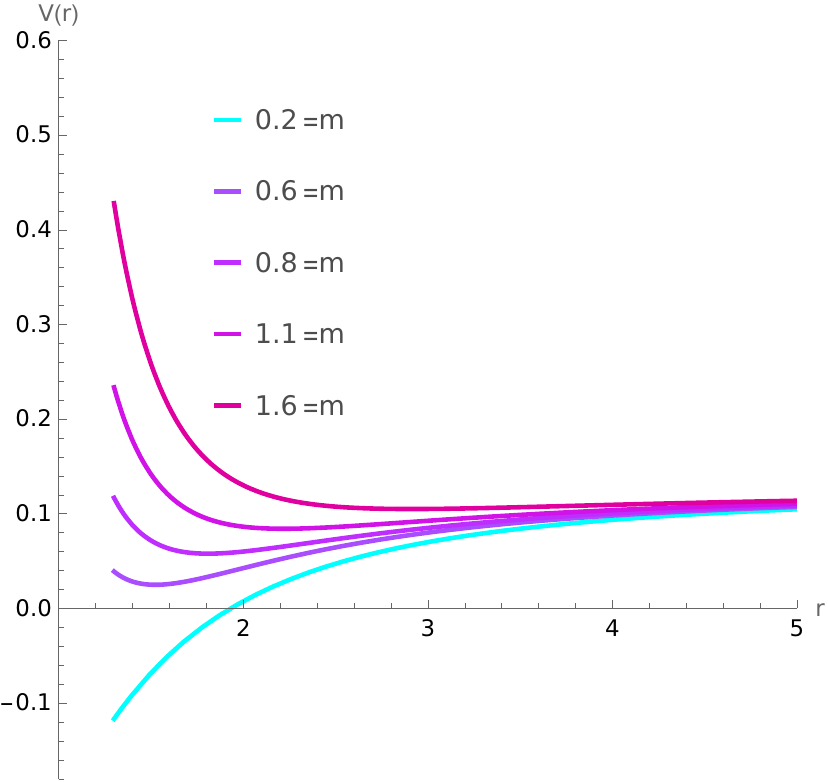}
			}
			\caption{Schr\"{o}dinger potential for longitudinal perturbations of the Wilson loop in background $\mathrm{IV}$ as $r_0$  and $m$ vary.}
			\label{VIV}
\end{figure}

\begin{figure}[H]
	\centering
	\includegraphics[width=0.75\textwidth]{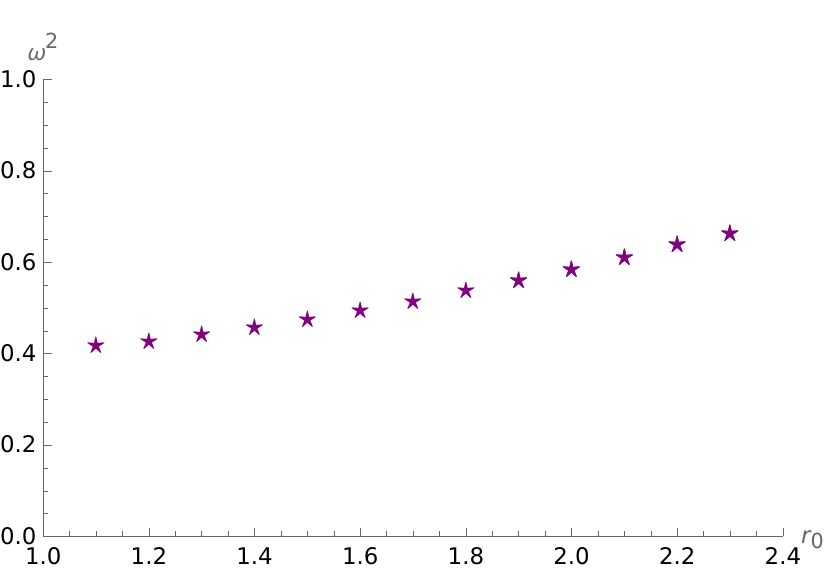}
	\caption{Plot of the lowest eigenvalues of the Schr\"{o}dinger problem for background IV as a function of $r_0$ for $m=0.2$.}
	\label{EigenIV}
\end{figure}

The plot of the potential can be seen in Figure \ref{VIV} and the numerical solution for the lowest eigenvalues in \ref{EigenIV}, where we deduce that the embedding is stable under longitudinal fluctuations. 

\subsubsection{'t Hooft loop}
We now move to the 't Hooft loop for this background, for which we embed a probe $\mathrm{D}5$ brane in the submanifold spanned by $[t,r,\mu,\theta,\phi,\psi]$, yielding the following effective action after performing the integrals:

\begin{equation}
	S ^{(\mathrm{IV})} _{\mathrm{D}5}= \mathrm{T} L _{\mu}16\pi^2 \int _{r_0}^{\infty}\mathrm{d}r \sqrt{f _{x_1}(r) x ^{\prime 2}+ g(r)},
\end{equation}

where $\mathrm{T}$ is the temporal length of the loop and the functions $f _{x_1}$ and $g$ read in this case:

\begin{equation}
	\begin{split}
		&f^{(\mathrm{IV})} _{t}(r) = r^2 (r^2-m),\\
		&g^{(\mathrm{IV})} _{t}(r) = 2r^2,
	\end{split}
\end{equation}

where it is evident that the monopole-anti-monopole string tension vanishes at the end of the space: $f ^{(\mathrm{IV})} _{t}(\sqrt{m})=0$ and therefore we expect this embedding to be unstable, much like the other cases covered. The length and its approximate expression for this system are:

\begin{equation}
	L ^{(\mathrm{IV})} _{\mathrm{MM}}(r_0) = \sqrt{2}r_0 \sqrt{r_0^2-m}\int _{r_0}^{\infty} \frac{\mathrm{d}r}{\sqrt{(r^2-m)(r^2-r_0^2)(r^2+r_0^2-m)}},
\end{equation}

\begin{equation}
	\hat{L}^{(\mathrm{IV})} _{\mathrm{MM}}(r_0) = \frac{\sqrt{2}\pi r \sqrt{r_0^2-m}}{2r_0^2-m}, 
\end{equation}

where we can easily read that the last expression approaches zero as $r_0\to \sqrt{m}$. The fluctuations obey:

\begin{equation}\label{VIVth}
	- \frac{\mathrm{d}}{\mathrm{d}r}\left[ P _{\mathrm{MM}} ^{(\mathrm{IV})}(r;r_0) \frac{\mathrm{d}}{\mathrm{d}r}  \right] \delta x_1(r) = \omega^2 W_{\mathrm{MM}} ^{(\mathrm{IV})}(r;r_0)\delta x_1 (r) ,
\end{equation}

where:

\begin{equation}
	\begin{split}
		& P_{\mathrm{MM}} ^{(\mathrm{IV})}(r;r_0) =\frac{\left[ (r^2-r_0^2) \left( r^2+r_0^2-m \right)   \right]^{3/2} }{r^2\sqrt{2(r^2-m)}},\\
		&W_{\mathrm{MM}} ^{(\mathrm{IV})}(r;r_0) =\frac{\sqrt{2(r^2-r_0^2)(r^2+r_0^2-m)}}{\sqrt{r^2-m}}.
	\end{split}
\end{equation}

\begin{figure}[H]
	\centering
	\subfloat[$m=1$]{
		\includegraphics[width=65mm]{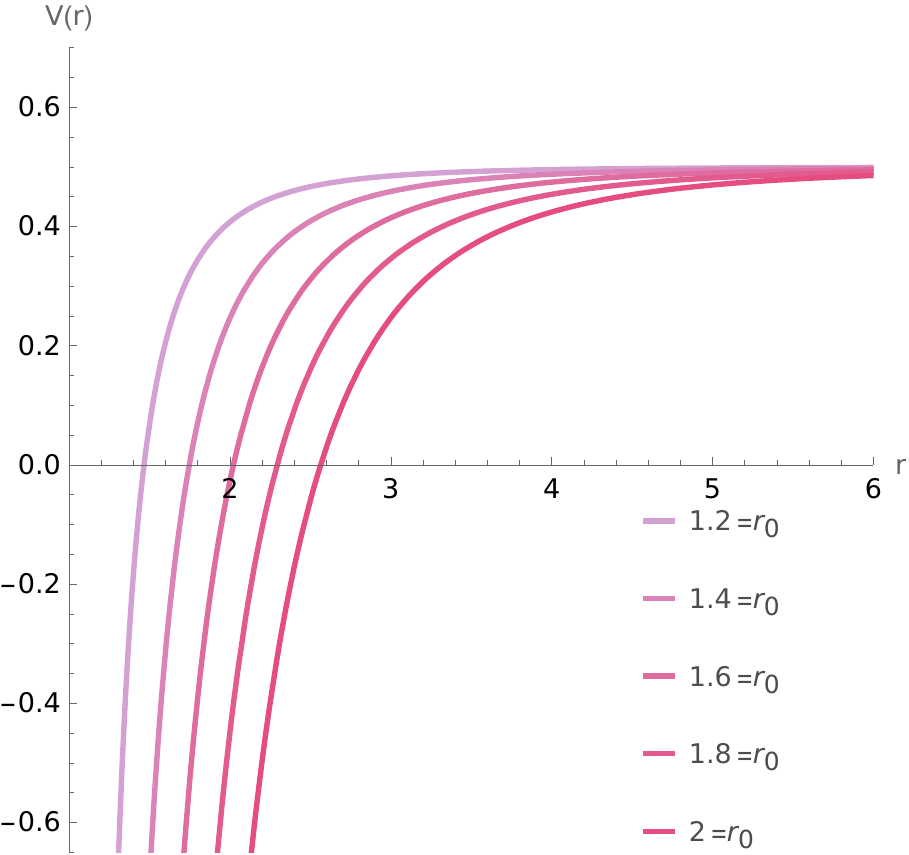}
		}
		\subfloat[$r _0=1.2$]{
			\includegraphics[width=65mm]{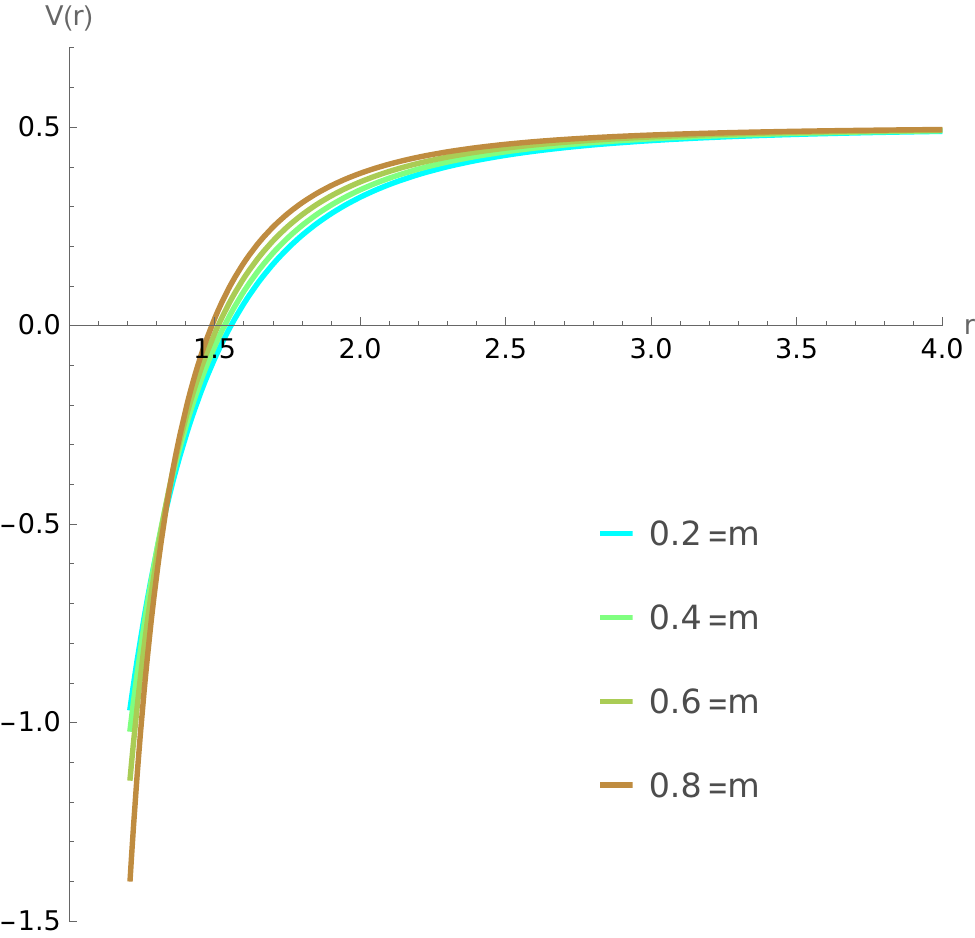}
			}
			\caption{Schr\"{o}dinger potential for longitudinal perturbations of the 't Hooft loop in background $\mathrm{IV}$ as $r_0$  and $m$ vary.}
			\label{VIVt}
\end{figure}

\begin{figure}[H]
	\centering
	\includegraphics[width=0.75\textwidth]{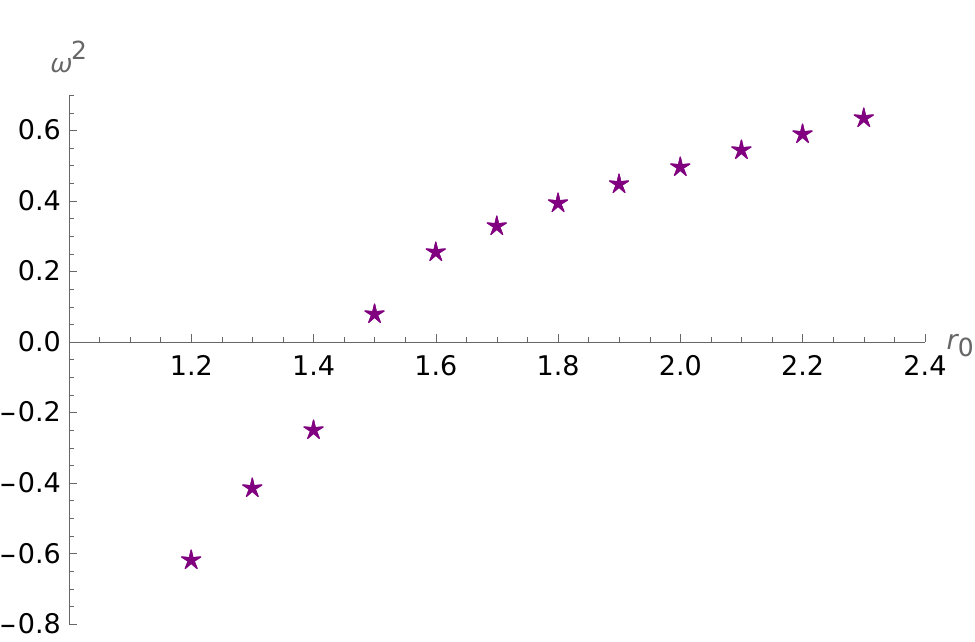}
	\caption{Plot of the lowest eigenvalues of the 't Hooft loop for background IV as a function of $r_0$ ($m=1$).}
	\label{VIVei}
\end{figure}

The potentials, which asymptote to $1/2$ for large $r$ values and have the same form as in the case of background $\mathrm{III}$, are ploted in Figure \ref{VIVt}. We also performed the numerical analysis producing similar results shown in Figure \ref{VIVei},  therefore proving that \eqref{VIVth} admits negative eigenvalues and the system screens.

\section{Conclusions}
Summarizing the contents presented in this paper:\\

In Sections \ref{setup},\ref{bgs} we considered four ten-dimensional Type $\mathrm{IIB}$ Supergravity solutions that have been studied in \cite{CMR1,CMR2}, which describe stacks $\mathrm{D}5$ branes that either intersect or wrap a compact manifold. These backgrounds have very interesting holographically dual theories that are confining and exhibit the following behaviour: At low energies there is some effective Quantum Field Theory that in all cases except background $\mathrm{IV}$ preserves some amount of supersymmetry. As the energy becomes larger, the dimension of the dual theory is increased by one, supersymmetry (where preserved) is enhanced and the system is eventually described, after performing an S-duality, by the holographic dual of an $\mathrm{NS}5$ brane setup that is Little String Theory. \\

Section \ref{observables} defines the Wilson loop, 't Hooft loop and Entanglement Entropy, as well as how they are calculated using holography in this setting. These are the observables whose embeddings are tested for instabilities.\\

In Section \ref{pert} we studied in detail the problem of small fluctuations around a classical solution, based on \cite{SAS07,SAS08,Silva}. Although the problem is described in general by a coupled Sturm-Liouville system of the longitudinal, transverse and angular perturbations, for the cases under consideration the only relevant perturbation is the longitudinal one which is in fact decoupled. We reviewed how it is more convenient to map the Sturm-Liouville equation into a Sch\"{o}dinger one and study the spectrum of the eigenvalues using the later. The existence of a negative eigenvalue mode implies an instability of the embedding under the corresponding perturbation.\\

We finally apply the stability analysis to the four Supergravity backgrounds in Section \ref{stabilitystudy}, studying the differential equations and ploting the Schr\"{o}dinger potentials for each case, in order to determine if there are any negative eigenvalues. This is done both by examining the plots and the length function, as well as using numerical methods where the eigenvalues are ploted with respect to the tip of the string.\\

The results found in the present work agree with the claims in \cite{CMR1,CMR2} where the stability was speculated using the concavity condition. In particular, the Wilson loop U-shaped embeddings were stable in all backgrounds while the 't Hooft loops and Entanglement entropy (where calculated) were found to be unstable under longitudinal fluctuations. Furthermore, our findings reflected the fact that the dual field theories are not UV completed by a field-theoretical system and also validated the general behaviour that when the Wilson loops confine the 't Hooft loops exhibit screening \cite{thooft1978}, which is evident as small fluctuations are found to destabilize the embedding into the disconnected one of a monopole and an anti-monopole moving freely.\\

There has been a new family of asymptotically AdS Type $\mathrm{IIB}$ backgrounds recently proposed in \cite{Fatemiabhari_2024}, that is achieved by uplifting a six-dimensional Romans $F _{4}$ gauged Supergravity \cite{Romans_1985}. These ten-dimensional backgrounds model a five-dimensional family of superconformal field theories, that is expressed using balanced linear quivers. It was noted that even though one can compute the Wilson loop for a gauge node with no flavour group attached, much in the same way as we have described here, there are still massless flavour quarks at some node in the quiver that need to be taken into account. The presence of dynamical flavour quarks is crucial in the study of confinement, as these quarks can break the flux tube and cause screening. This is not captured by the usual embedding in the $(x_1(\sigma),r(\sigma))$ plane, but one needs to probe the string in the $\eta=\eta(\sigma)$ direction that labels the gauge nodes of the quiver as well. It will be interesting in the future to apply the stability analysis for embeddings in this setting, and in particular investigate the role of fluctuations around the $\eta$ direction.

\section*{Acknowledgements}
The author would like to thank Carlos Nu$\tilde{\mathrm{n}}$ez for his guidance, support and continuous conversations on topics relevant but not limited to holography and string theory. Conversations with Ricardo Stuardo and Ali Fatemiabhari were also vital for understanding much of the background needed for this work, while a special gratitude is expressed to A.F. for the help he provided with the shooting method code. This work is supported by STFC DTP grant No. ST/Y509644-1. For the purpose of open access, the author has applied a Creative Commons Attribution (CC BY) licence to any Author Accepted Manuscript version arising.

\appendix
\counterwithin*{equation}{section}
\renewcommand\theequation{\thesection\arabic{equation}}

\section{Zero modes of longitudinal perturbations $\Leftrightarrow L^{\prime}(r_0)=0$}\label{zeromodes}

Up to a constant, the solution for the zero modes of the equation of motion for the longitudinal fluctuation, namely\footnote{We make use of the abbreviations: $g(r)\equiv g,f _{x_1}(r)\equiv ,f _{x_1},f _{x_1}(r_0) \equiv f _{x_10}$.}, 

\begin{equation}
	\begin{split}
		&\frac{\mathrm{d}}{\mathrm{d}r} \left( \frac{f _{x_1}g}{F ^{3/2}}\delta x_1 ^{\prime (0)}(r) \right) =0  \Rightarrow \delta x ^{(0)} _1(r) = \int _{r}^{\infty}\mathrm{d}\xi \frac{F ^{3/2}(\xi)}{f _{x_1}(\xi)g(\xi)}\\
		& \Rightarrow \delta x ^{(0)} _1 (r) = \int _{r}^{\infty}\mathrm{d}\xi \frac{\sqrt{f _{x_1}g}}{\left(f _{x_1} - f _{x_10}  \right)^{3/2} },
	\end{split}
\end{equation}

where we understand everything inside the integral as a function of $\xi$. We can now use the following identity:

\begin{equation}\label{iden}
	(f _{x_1}-f _{x_10})^{-3/2} = - \frac{2}{f ^{\prime} _{x_1}}\frac{\partial}{\partial \xi } \left( f _{x_1} - f _{x_10} \right) ^{-1/2}
\end{equation}
and perform integration by parts to get another expression for $\delta x ^{(0)}$:

\begin{equation}
	\begin{split}
		&		\delta x ^{(0)}(r) = - \int _r ^{\infty}\mathrm{d}\xi \sqrt{f _{x_1}g} \frac{2}{f ^{\prime} _{x_1}} \frac{\partial }{\partial \xi} \left( \frac{1}{\sqrt{f _{x_1}- f _{x_10}}} \right) \\
		&= - \left. \frac{2}{f ^{\prime}_{x_1}} \sqrt{ \frac{f _{x_1}g}{f _{x_1}- f _{x_10}}} \right|^{\infty}_{r} +2 \int _r ^{\infty} \frac{\mathrm{d}\xi}{\sqrt{f _{x_1}-f _{x_10}}} \frac{\partial }{\partial \xi} \left( \frac{\sqrt{f _{x_1}g}}{f ^{\prime}_{x_1}} \right) \\
		&=  \lim _{\xi\to r} \frac{2}{f ^{\prime}_{x_1}} \sqrt{ \frac{f _{x_1}g}{f _{x_1}- f _{x_10}}} +2 \int _r ^{\infty} \frac{\mathrm{d}\xi}{\sqrt{f _{x_1}-f _{x_10}}} \frac{\partial }{\partial \xi} \left( \frac{\sqrt{f _{x_1}g}}{f ^{\prime}_{x_1}} \right) \\
		& = 2 \sqrt{\frac{f _{x_10}g(r_0)}{f ^{\prime 3}_{x_10}}}(r-r_0)^{-1/2}+2 \int _r ^{\infty} \frac{\mathrm{d}\xi}{\sqrt{f _{x_1}-f _{x_10}}} \frac{\partial }{\partial \xi} \left( \frac{\sqrt{f _{x_1}g}}{f ^{\prime}_{x_1}} \right),
	\end{split}
\end{equation}

where in the third line we used that the first term vanishes at $r\to \infty$. In order for the zero mode to be a solution of the problem, we need to impose the boundary condition \eqref{BCdx1} for $r\to r_0$, namely:

\begin{equation}
	\delta x_1 ^{{0}}(r) + 2(r-r_0)\delta x ^{\prime(0)}_1 (r) = 0,
\end{equation}

\begin{equation}
	\begin{split}
	\Rightarrow	&2 \sqrt{ \frac{f _{x_10}g(r_0)}{f^{\prime 3}_{x_10}}}(r-r_0)^{-1/2} + 2 \int _{r_0}^{\infty}\frac{\mathrm{d}\xi}{\sqrt{f _{x_1}-f_{x_10}}}\frac{\partial}{\partial \xi}\left( \frac{\sqrt{f _{x_1}g}}{f^{\prime}_{x_1}} \right) \\
		&+2(r-r_0)\left[-\sqrt{ \frac{f _{x_10}g(r_0)}{f^{\prime 3}_{x_10}}}(r-r_0)^{-3/2}  + \frac{2}{\sqrt{f _{x_1}-f _{x_10}}} \frac{\partial}{\partial r}\left( \frac{\sqrt{f _{x_1}g}}{f ^{\prime} _{x_1}} \right)  \right]=0,
	\end{split}
\end{equation}

\begin{equation}
	\Rightarrow  \frac{4(r-r_0)}{\sqrt{f _{x_1}-f _{x_10}}}\frac{\partial}{\partial r}\left( \frac{\sqrt{f _{x_1}g}}{f ^{\prime} _{x_1}} \right)  + 2 \int _{r_0}^{\infty}\frac{\mathrm{d}\xi}{\sqrt{f _{x_1}-f_{x_10}}}\frac{\partial}{\partial \xi}\left( \frac{\sqrt{f _{x_1}g}}{f^{\prime}_{x_1}} \right)  =0.
\end{equation}

As $r\to r_0$, the first term goes like $\sim \sqrt{r-r_0}$ so it vanishes.  This leaves us with the following condition in order for a longitudinal zero mode to exist:

\begin{equation}\label{condition}
\int _{r_0}^{\infty}\frac{\mathrm{d}\xi}{\sqrt{f _{x_1}-f_{x_10}}}\frac{\partial}{\partial \xi}\left( \frac{\sqrt{f _{x_1}g}}{f^{\prime}_{x_1}} \right)  =0.
\end{equation}

Let us now calculate the derivative of the length from \eqref{length}:

\begin{equation}
	L^{\prime}(r_0) = 2 \int _{r_0}^{\infty}\mathrm{d}r \left[ \frac{f^{\prime}_{x_10}}{2 \sqrt{f _{x_10}}}\frac{\sqrt{F}}{f _{x_1}} + \frac{\sqrt{f _{x_10}}}{2f _{x_1}\sqrt{F}} \frac{\partial F}{\partial r_0}  \right] + 2 \left.\frac{\sqrt{f _{x_10}F}}{f _{x_1}}\right|^{\infty} _{r_0} 
\end{equation}

\begin{equation}
	= \frac{f ^{\prime} _{x_10}}{\sqrt{ f _{x_10}}}\int _{r_0}^{\infty}\mathrm{d}r \frac{\sqrt{f _{x_1}g}}{\left( f _{x_1} -f _{x_10} \right) ^{3/2}} -2 \lim _{r\to r_0} \frac{\sqrt{f _{x_10}F}}{f _{x_1}},
\end{equation}

again, using \eqref{iden},

\begin{equation}
	\begin{split}
		=& - \left.\frac{2 f^{\prime} _{x_10}}{\sqrt{f _{x_10}}}\frac{\sqrt{f _{x_1}g}}{f ^{\prime}_{x_1}} \frac{1}{\sqrt{f _{x_1} - f _{x_10}}}\right| ^{\infty} _{r_0} + \frac{2 f^{\prime} _{x_10}}{\sqrt{f _{x_10}}}\int _{r_0}^{\infty} \frac{\mathrm{d}r}{\sqrt{f _{x_1} - f _{x_10}}}\frac{\partial}{\partial r} \left( \frac{\sqrt{f _{x_1}g}}{f ^{\prime} _{x_1}} \right) \\
		&-2 \lim _{r\to r_0} \frac{\sqrt{f _{x_10}F}}{f _{x_1}} \frac{f _{x_10}}{f _{x_1}},
	\end{split}
\end{equation}

\begin{equation}
	\begin{split}
		=&  2\lim _{r\to r_0}\frac{ \sqrt{F}}{\sqrt{f _{x_10}}} \frac{f ^{\prime} _{x_10}}{f ^{\prime} _{x_1}}+ \frac{2 f^{\prime} _{x_10}}{\sqrt{f _{x_10}}}\int _{r_0}^{\infty} \frac{\mathrm{d}r}{\sqrt{f _{x_1} - f _{x_10}}}\frac{\partial}{\partial r} \left( \frac{\sqrt{f _{x_1}g}}{f ^{\prime} _{x_1}} \right) \\
		&-2 \lim _{r\to r_0} \frac{\sqrt{f _{x_10}F}}{f _{x_1}} \frac{f _{x_10}}{f _{x_1}},
	\end{split}
\end{equation}
where we multiplied and devided by the last factor on the third term, which is 1 inside the limit. We can then join the two limits which cancel each other, to get:

\begin{equation}
=2 \lim _{r\to r_0} \frac{\sqrt{F}}{\sqrt{f _{x_10}}} \left( \frac{f ^{\prime} _{x_10}}{f ^{\prime} _{x_1}} - \frac{f _{x_1}}{f _{x_10}} \right)+  \frac{2 f^{\prime} _{x_10}}{\sqrt{f _{x_10}}}\int _{r_0}^{\infty} \frac{\mathrm{d}r}{\sqrt{f _{x_1} - f _{x_10}}}\frac{\partial}{\partial r} \left( \frac{\sqrt{f _{x_1}g}}{f ^{\prime} _{x_1}} \right) 
\end{equation}

\begin{equation}\label{L'}
	\Rightarrow L ^{\prime} (r_0) =   \frac{2 f^{\prime} _{x_10}}{\sqrt{f _{x_10}}}\int _{r_0}^{\infty} \frac{\mathrm{d}r}{\sqrt{f _{x_1} - f _{x_10}}}\frac{\partial}{\partial r} \left( \frac{\sqrt{f _{x_1}g}}{f ^{\prime} _{x_1}} \right) .
\end{equation}

Thus, the condition \eqref{condition} can be written as:

\begin{equation}
	\frac{\sqrt{f _{x_1}}}{ f^{\prime} _{x_10}} L ^{\prime} (r_0) = 0,
\end{equation}

from which we are lead to the conclusion that the zero modes of the longitudinal perturbations are in 1-1 correspondance with the critical points of the length. This informs us that for our examples in this paper, where the length functions are all monotonous, there are no zero eigenvalue modes. One more fact to note is that due to \eqref{dEdL}, namely,

\begin{equation}
	E ^{\prime}(r_0) = \frac{\sqrt{f _{x0}}}{2\pi}L ^{\prime}(r_0)
\end{equation}

we see that the critical points of the length and the energy coincide.

\section{Schr\"{o}dinger description of the problem}\label{SCH}
Here we present the change of variables that maps the Sturm-Liouville problem for the fluctuations to a Schr\"{o}dinger equation in some more detail \cite{SAS07,Silva}. The general form of the Sturm-Liouville equation satisfied by any decoupled fluctuation $\delta\Phi$ is:

\begin{equation}
	- \frac{\mathrm{d}}{\mathrm{d}r}\left[ P(r;r_0) \frac{\mathrm{d}}{\mathrm{d}r} + Q(r;r_0)  \right] \delta \Phi (r) = \omega ^2 W(r;r_0) \delta\Phi(r) \,\, , r ^{\star} \leq r_0\leq r<\infty,
\end{equation}

where $Q\neq 0$ expresses the restoring force term originating from possible dependence of the metric components on $\Phi$, if $\Phi$ is an angular coordinate (which in the cases we are going to study is zero). The only $"$angular$"$ dependence of the metric components is seen in backgrounds $\mathrm{II}$ and $\mathrm{III}$ but it does not produce such a restoring force term in the equations of motion. Using the following coordinate transformation, this Sturm-Liouville problem can be translated to a Schr\"{o}dinger equation:\\
\begin{equation}\label{cov}
	y(r;r_0):= \int _{r_0}^{r} \mathrm{d}\xi \sqrt{ \frac{W (\xi,r_0)}{P (\xi,r_0)}} = \int _{r_0}^{r}\mathrm{d}\xi \sqrt{\frac{h(\xi)}{f _{x_1}(\xi)-f _{x_1}(r_0)}},
\end{equation}

Then the Schr\"{o}dinger wavefunction satisfying the equation is:

\begin{equation}\label{Psi}
	\Psi (y) := \left[ P (r;r_0) W (r;r_0)  \right]^{1/4}\delta\Phi(r), 
\end{equation}

and \eqref{SL} is turned into:

\begin{equation}\label{ssss}
	\left[ - \frac{\mathrm{d}^2}{\mathrm{d}y^2} + V(y,r_0)  \right] \Psi (y) = \omega ^2 \Psi (y)\,\, , 0\leq y \leq y_0  ,
\end{equation}
where $y_0$ is defined as:

\begin{equation}\label{changeofvar}
	y_0(r_0)= \int _{r_0} ^{\infty}\mathrm{d}\xi \sqrt{ \frac{h(\xi)}{f _{x_1}(\xi) - f _{x_1}(r_0)}}.
\end{equation}

If the space has an asymptotically AdS factor $y_0$ is finite in both integration limits, but in general it can be either finite or infinite depending on the details of the geometry. The  interval $[r_0,\infty)$ is mapped into $[0,y_0]$ under \eqref{cov} and the boundary conditions for \eqref{SE} are: 

\begin{equation}
\begin{split}
	&\Psi(y_0)=0\,\, , \,\,\text{from the fixed points of the string at} \,r=\infty\\
	&\frac{\mathrm{d}\Psi}{\mathrm{d}y}\Big| _{y_0}=0\,\, , \,\, \text{from the condition at the tip}
\end{split}
\end{equation}

The Schr\"{o}dinger potential in \eqref{ssss} is then found to be:

\begin{equation}\label{Vyap}
	V(y;r_0) = - \frac{Q }{W } + \left( P W  \right) ^{-1/4} \frac{\mathrm{d}^2}{\mathrm{d}y^2} \left( P W  \right) ^{1/4},
\end{equation}

or, expressed in terms of $r$:

\begin{equation}
		V (r,r_0) = - \frac{Q }{W } + \frac{P^{1/4} }{W ^{3/4} } \frac{\mathrm{d}}{\mathrm{d}r} \left[ \left( \frac{P }{W } \right) ^{1/2} \frac{\mathrm{d}}{\mathrm{d}r}\left( P W  \right) ^{1/4}  \right] .
\end{equation}
Although it is not often that one can invert $y=y(r)$ in order to get \eqref{Vyap}, the expression in terms of $r$ which follows from the equations of motion can provide enough evidence for the study of the spectrum.
We also note that for all Wilson loops of backgrounds $(\mathrm{I}-\mathrm{IV})$ we have the asymptotic behaviour: As $r\to \infty$, $\Psi (y) \sim \sqrt{r}\delta \Phi(r)$, while for the 't Hooft loops and Entanglement Entropy: $\Psi (y) \sim r \delta \Phi(r)$.

\printbibliography[title=References]

\typeout{get arXiv to do 4 passes: Label(s) may have changed. Rerun}

\end{document}